\documentclass[a4paper,11pt]{article}

\usepackage{amsfonts,amsmath,amssymb}
\usepackage{yhmath}
\usepackage{physics}
\usepackage{tensor}
\usepackage{array}
\usepackage{mathtools}
\usepackage{braket}
\usepackage{slashed}
\usepackage{cancel}
\usepackage{leftidx}
\usepackage{stackrel}

\newcommand{\mC}{\mathcal{C}}

\newcommand{\mN}{\mathcal{N}}
\newcommand{\mD}{\mathcal{D}}
\newcommand{\mB}{\mathcal{B}}
\newcommand{\mP}{\mathcal{P}}
\newcommand{\mZ}{\mathcal{Z}}

\newcommand{\vareps}{\varepsilon}

\newcommand{\Cpp}{C_{\phi\phi}}

\newcommand{\no}{\nonumber}
\newcommand{\der}{\partial}
\newcommand{\be}{\begin{equation}}
\newcommand{\ee}{\end{equation}}

\newcommand{\Vol}{\operatorname{Vol}}

\usepackage{enumerate}
\usepackage{bm}
\usepackage[thinlines]{easytable} 
\usepackage{tabu}
\usepackage[normalem]{ulem}

\usepackage{xcolor}
\colorlet{darkblue}{blue!70!black}
\usepackage[colorlinks=true,urlcolor=darkblue,linktocpage=true,linkcolor=darkblue,citecolor=green]{hyperref}

\newcommand{\arxiv}[1]{arXiv:\href{http://www.arXiv.org/abs/#1}{#1}}

\usepackage{graphicx,epstopdf}
\usepackage{subcaption}

\usepackage{tikz}
\usepackage[compat=1.1.0]{tikz-feynman}

\graphicspath{{./figures/}}

\newcommand{\proponeloop}{
\begin{tikzpicture}
  \begin{feynman}
 \vertex (i1) at (0,0)  {\(\psi\)};
 \vertex (i2) at (3,0)  {\(\psi\)};
 \vertex (a) at (1,0);  
 \vertex (b) at (2,0);
\diagram* {
(i1) -- [fermion] (a) -- [fermion] (b) -- [fermion] (i2),
(a) -- [scalar, half left, looseness=1.6] (b)
};
\end{feynman}
\end{tikzpicture}
}

\newcommand{\proponeloopCT}{
\begin{tikzpicture}
  \begin{feynman}
 \vertex (i1) at (0,0)  {\(\psi\)};
 \vertex (i2) at (3,0)  {\(\psi\)};
 \vertex (a) at (1.5,0);  
\diagram* {
(i1) -- [fermion] (a) -- [fermion] (i2),
};
 \node at (a) {\(\times\)};
\end{feynman}
\end{tikzpicture}
}

\newcommand{\vertexoneloop}{
\begin{tikzpicture}
  \begin{feynman}
 \vertex (i1) at (0,0) {\(\psi\)};
 \vertex (i2) at (4,0) {\(\psi\)};
 \vertex (a) at (1,0);
 \vertex (b) at (2,0);
 \vertex (c) at (3,0);  
 \vertex (f1) at (2,1) {\(\phi\)};
    \diagram*  {
      (i1) -- [fermion] (a) -- [fermion] (b) -- [fermion] (c) -- [fermion] (i2) ,
      (a) -- [scalar, half right, looseness=1.6] (c),
      (f1) -- [scalar] (b)
       };
         \end{feynman}
\end{tikzpicture}
}

\newcommand{\vertexoneloopCT}{
\begin{tikzpicture}
  \begin{feynman}
 \vertex (i1) at (0,0)  {\(\psi\)};
 \vertex (i2) at (3,0)  {\(\psi\)};
 \vertex (a) at (1.5,0);  
 \vertex (f1) at (1.5,1) {\(\phi\)};
    \diagram*  {
      (i1)--  [fermion] (a) -- [fermion] (i2) ,
       (f1)-- [scalar] (a)
       };
        \node at (a) {\(\times\)};
         \end{feynman}
\end{tikzpicture}
}

\newcommand{\proptwoloopone}{
\begin{tikzpicture}
  \begin{feynman}
 \vertex (i1) at (0,0) {\(\psi\)};
 \vertex (i2) at (5,0) {\(\psi\)};
 \vertex (a) at (1,0);
 \vertex (b) at (2,0);
 \vertex (c) at (3,0);  
 \vertex (d) at (4,0);
\diagram* {
 (i1) -- [fermion] (a) -- [fermion] (b) -- [fermion] (c) -- [fermion] (d)  -- [fermion] (i2),
      (a) -- [scalar, half left, looseness=1.4] (d),
      (b) -- [scalar, half left, looseness=1.6] (c),
};
\end{feynman}
\end{tikzpicture}
}

\newcommand{\proptwolooptwo}{
\begin{tikzpicture}
  \begin{feynman}
 \vertex (i1) at (0,0) {\(\psi\)};
 \vertex (i2) at (5,0) {\(\psi\)};
 \vertex (a) at (1,0);
 \vertex (b) at (2,0);
 \vertex (c) at (3,0);  
 \vertex (d) at (4,0);
\diagram* {
 (i1) -- [fermion] (a) -- [fermion] (b) -- [fermion] (c) -- [fermion] (d)  -- [fermion] (i2),
      (a) -- [scalar, half right, looseness=1.6] (c),
      (b) -- [scalar, half left, looseness=1.6] (d),
};
\end{feynman}
\end{tikzpicture}
}

\newcommand{\proptwolooponeCT}{
\begin{tikzpicture}
  \begin{feynman}
 \vertex (i1) at (0,0) {\(\psi\)};
 \vertex (i2) at (4,0) {\(\psi\)};
 \vertex (a) at (1,0);
 \vertex (b) at (2,0);
 \vertex (c) at (3,0);  
\diagram* {
 (i1) -- [fermion] (a) -- [fermion] (b) -- [fermion] (c) --  [fermion] (i2),
      (a) -- [scalar, half left, looseness=1.4] (c)
   };
  \node at (b) {\(\times\)};
\end{feynman}
\end{tikzpicture}
}

\newcommand{\proptwolooptwoCTa}{
\begin{tikzpicture}
  \begin{feynman}
 \vertex (i1) at (0,0) {\(\psi\)};
 \vertex (i2) at (4,0) {\(\psi\)};
 \vertex (a) at (1,0);
 \vertex (c) at (3,0);  
\diagram* {
 (i1) -- [fermion] (a) -- [fermion] (c)  -- [fermion] (i2),
      (a) -- [scalar, half right, looseness=1.6] (c),
};
 \node at (c) {\(\times\)};
\end{feynman}
\end{tikzpicture}
}

\newcommand{\proptwolooptwoCTb}{
\begin{tikzpicture}
  \begin{feynman}
 \vertex (i1) at (0,0) {\(\psi\)};
 \vertex (i2) at (4,0) {\(\psi\)};
 \vertex (a) at (1,0);
 \vertex (c) at (3,0);  
\diagram* {
 (i1) -- [fermion] (a) -- [fermion] (c)  -- [fermion] (i2),
      (a) -- [scalar, half left, looseness=1.6] (c),
};
 \node at (a) {\(\times\)};
\end{feynman}
\end{tikzpicture}
}

\newcommand{\vertextwoloopone}{
\begin{tikzpicture}
  \begin{feynman}
 \vertex (i1) at (0,0) {\(\psi\)};
 \vertex (i2) at (6,0) {\(\psi\)};
 \vertex (a) at (1,0);
 \vertex (b) at (2,0);
 \vertex (c) at (3,0);  
  \vertex (d) at (4,0);  
    \vertex (e) at (5,0);  
 \vertex (f1) at (3,1.5) {\(\phi\)};
    \diagram*  {
      (i1) [particle=\(e^-\)] -- [fermion] (a) -- [fermion] (b) -- [fermion] (c) -- [fermion] (d) -- [fermion] (e) -- [fermion] (i2) [particle=\(e^+\)],
      (a) -- [scalar, half right, looseness=1.4] (e),
      (b) -- [scalar, half right, looseness=1.6] (d),
      (f1) -- [scalar] (c)
       };
       \end{feynman}
\end{tikzpicture}
}

\newcommand{\vertextwolooptwo}{
\begin{tikzpicture}
  \begin{feynman}
 \vertex (i1) at (0,0) {\(\psi\)};
 \vertex (i2) at (6,0) {\(\psi\)};
 \vertex (a) at (1,0);
 \vertex (b) at (2,0);
 \vertex (c) at (3,0);  
  \vertex (d) at (4,0);  
    \vertex (e) at (5,0);  
 \vertex (f1) at (3,1.5) {\(\phi\)};
    \diagram*  {
      (i1) [particle=\(e^-\)] -- [fermion] (a) -- [fermion] (b) -- [fermion] (c) -- [fermion] (d) -- [fermion] (e) -- [fermion] (i2) [particle=\(e^+\)],
      (a) -- [scalar, half right, looseness=1.6] (d),
      (b) -- [scalar, half right, looseness=1.6] (e),
      (f1) -- [scalar] (c)
       };
       \end{feynman}
\end{tikzpicture}
}

\newcommand{\vertextwoloopthree}{
\begin{tikzpicture}
  \begin{feynman}
 \vertex (i1) at (0,0) {\(\psi\)};
 \vertex (i2) at (6,0) {\(\psi\)};
 \vertex (a) at (1,0);
 \vertex (b) at (2,0);
 \vertex (c) at (3,0);  
  \vertex (d) at (4,0);  
    \vertex (e) at (5,0);  
 \vertex (f1) at (4,1.5) {\(\phi\)};
    \diagram*  {
      (i1) [particle=\(e^-\)] -- [fermion] (a) -- [fermion] (b) -- [fermion] (c) -- [fermion] (d) -- [fermion] (e) -- [fermion] (i2) [particle=\(e^+\)],
      (a) -- [scalar, half left, looseness=1.4] (c),
      (b) -- [scalar, half right, looseness=1.6] (e),
      (f1) -- [scalar] (d)
       };
       \end{feynman}
\end{tikzpicture}
}

\newcommand{\vertextwoloopfour}{
\begin{tikzpicture}
  \begin{feynman}
 \vertex (i1) at (0,0) {\(\psi\)};
 \vertex (i2) at (6,0) {\(\psi\)};
 \vertex (a) at (1,0);
 \vertex (b) at (2,0);
 \vertex (c) at (3,0);  
  \vertex (d) at (4,0);  
    \vertex (e) at (5,0);  
 \vertex (f1) at (2,1.5) {\(\phi\)};
    \diagram*  {
      (i1) [particle=\(e^-\)] -- [fermion] (a) -- [fermion] (b) -- [fermion] (c) -- [fermion] (d) -- [fermion] (e) -- [fermion] (i2) [particle=\(e^+\)],
      (c) -- [scalar, half left, looseness=1.4] (e),
      (a) -- [scalar, half right, looseness=1.6] (d),
      (f1) -- [scalar] (b)
       };
       \end{feynman}
\end{tikzpicture}
}

\newcommand{\vertextwoloopfive}{
\begin{tikzpicture}
  \begin{feynman}
 \vertex (i1) at (0,0) {\(\psi\)};
 \vertex (i2) at (6,0) {\(\psi\)};
 \vertex (a) at (1,0);
 \vertex (b) at (2,0);
 \vertex (c) at (3,0);  
  \vertex (d) at (4,0);  
    \vertex (e) at (5,0);  
 \vertex (f1) at (4,1.5) {\(\phi\)};
    \diagram*  {
      (i1) [particle=\(e^-\)] -- [fermion] (a) -- [fermion] (b) -- [fermion] (c) -- [fermion] (d) -- [fermion] (e) -- [fermion] (i2) [particle=\(e^+\)],
      (a) -- [scalar, half right, looseness=1.2] (e),
      (b) -- [scalar, half left, looseness=1.6] (c),
      (f1) -- [scalar] (d)
       };
       \end{feynman}
\end{tikzpicture}
}

\newcommand{\vertextwoloopsix}{
\begin{tikzpicture}
  \begin{feynman}
 \vertex (i1) at (0,0) {\(\psi\)};
 \vertex (i2) at (6,0) {\(\psi\)};
 \vertex (a) at (1,0);
 \vertex (b) at (2,0);
 \vertex (c) at (3,0);  
  \vertex (d) at (4,0);  
    \vertex (e) at (5,0);  
 \vertex (f1) at (2,1.5) {\(\phi\)};
    \diagram*  {
      (i1) [particle=\(e^-\)] -- [fermion] (a) -- [fermion] (b) -- [fermion] (c) -- [fermion] (d) -- [fermion] (e) -- [fermion] (i2) [particle=\(e^+\)],
      (a) -- [scalar, half right, looseness=1.2] (e),
      (c) -- [scalar, half left, looseness=1.6] (d),
      (f1) -- [scalar] (b)
       };
       \end{feynman}
\end{tikzpicture}
}

\newcommand{\vertextwolooponeCT}{
\begin{tikzpicture}
  \begin{feynman}
 \vertex (i1) at (0,0) {\(\psi\)};
 \vertex (i2) at (4,0) {\(\psi\)};
 \vertex (a) at (1,0);
 \vertex (b) at (2,0);
 \vertex (c) at (3,0);  
 \vertex (f1) at (2,1) {\(\phi\)};
    \diagram*  {
      (i1) [particle=\(e^-\)] -- [fermion] (a) -- [fermion] (b) -- [fermion] (c) -- [fermion] (i2) [particle=\(e^+\)],
      (a) -- [scalar, half right, looseness=1.6] (c),
      (f1) [particle=\(\phi\)] -- [scalar] (b)
       };
       \node at (b) {\(\times\)};
  \end{feynman}
\end{tikzpicture}
}

\newcommand{\vertextwoloopthreeCT}{
\begin{tikzpicture}
  \begin{feynman}
 \vertex (i1) at (0,0) {\(\psi\)};
 \vertex (i2) at (4,0) {\(\psi\)};
 \vertex (a) at (1,0);
 \vertex (b) at (2,0);
 \vertex (c) at (3,0);  
 \vertex (f1) at (2,1) {\(\phi\)};
    \diagram*  {
      (i1) [particle=\(e^-\)] -- [fermion] (a) -- [fermion] (b) -- [fermion] (c) -- [fermion] (i2) [particle=\(e^+\)],
      (a) -- [scalar, half right, looseness=1.6] (c),
      (f1) [particle=\(\phi\)] -- [scalar] (b)
       };     
       \node at (a) {\(\times\)};
  \end{feynman}
\end{tikzpicture}
}

\newcommand{\vertextwoloopfourCT}{
\begin{tikzpicture}
  \begin{feynman}
 \vertex (i1) at (0,0) {\(\psi\)};
 \vertex (i2) at (4,0) {\(\psi\)};
 \vertex (a) at (1,0);
 \vertex (b) at (2,0);
 \vertex (c) at (3,0);  
 \vertex (f1) at (2,1) {\(\phi\)};  
    \diagram*  {
      (i1) [particle=\(e^-\)] -- [fermion] (a) -- [fermion] (b) -- [fermion] (c) -- [fermion] (i2) [particle=\(e^+\)],
      (a) -- [scalar, half right, looseness=1.6] (c),
      (f1) [particle=\(\phi\)] -- [scalar] (b)
       };      
       \node at (c) {\(\times\)};
  \end{feynman}
\end{tikzpicture}
}

\newcommand{\vertextwoloopfiveCT}{
\begin{tikzpicture}
  \begin{feynman}
 \vertex (i1) at (0,0)  {\(\psi\)};
 \vertex (i2) at (4,0)  {\(\psi\)};
 \vertex (a) at (1,0);
 \vertex (b) at (2,0);
 \vertex (ab) at (1.5,0);
 \vertex (c) at (3,0);  
 \vertex (f1) at (2,1)  {\(\phi\)};
    \diagram*  {
      (i1) -- [fermion] (a) -- [fermion] (ab) -- [fermion] (b) -- [fermion] (c) -- [fermion] (i2),
      (a) -- [scalar, half right, looseness=1.6] (c),
      (f1)-- [scalar] (b)
       };
       \node at (ab) {\(\times\)};
  \end{feynman}
\end{tikzpicture}
}

\newcommand{\vertextwoloopsixCT}{
\begin{tikzpicture}
  \begin{feynman}
 \vertex (i1) at (0,0)  {\(\psi\)};
 \vertex (i2) at (4,0)  {\(\psi\)};
 \vertex (a) at (1,0);
 \vertex (b) at (2,0);
 \vertex (bc) at (2.5,0);
 \vertex (c) at (3,0);  
 \vertex (f1) at (2,1)  {\(\phi\)};
    \diagram*  {
      (i1) -- [fermion] (a) -- [fermion] (b) -- [fermion] (bc) -- [fermion] (c) -- [fermion] (i2),
      (a) -- [scalar, half right, looseness=1.6] (c),
      (f1)-- [scalar] (b)
       };
       \node at (bc) {\(\times\)};
  \end{feynman}
\end{tikzpicture}
}

\begin{document}
\title{How to get an interacting conformal line defect for free theories}
\author{Samuel Bartlett-Tisdall$^{\ddag}$, Dongsheng Ge$^{\dagger}$, and Christopher P. Herzog$^{\ddag}$}
\date{\today}

\maketitle

\vspace*{-6.3cm}
\begin{flushright}
OU-HET-1290
\end{flushright}
\vspace*{0.2cm}

\vspace*{5cm}
\centerline{$^\ddag$\it  Department of Mathematics, King’s College London,}
\centerline{\it The Strand WC2R 2LS, England}
\centerline{$^\dagger$\it Department of Physics, The University of Osaka,}
\centerline{\it Machikaneyama-Cho 1-1, Toyonaka 560-0043, Japan}


\vskip 1cm 


\begin{abstract}

\end{abstract}
We argue that interacting conformal line defects in free quantum field theories can exist, provided that inversion symmetry is broken.  
Important for our demonstration is the existence of a special cross ratio for bulk-defect-defect three point functions that is invariant under the conformal group but picks up a sign under inversion.
We examine the particular case of a free scalar field in detail, and  provide a toy model example where this bulk field interacts via a Yukawa term with fermions on the line.  We expect nontrivial line defects may also exist for free Maxwell theory in four dimensions and free bulk fermions.

\thispagestyle{empty}
\newpage
\tableofcontents



\section{Introduction}

While quantum field theories (QFTs) are conventionally tailored to the description of point particles
in flat space, defects in QFTs allow us on the one hand to model important physical effects and on 
the other to push QFTs into new regimes where the analytical tools at our disposal enjoy new
and unexplored power.  A field theoretic model of graphene for example should constrain the 
charged degrees of freedom to a $p = 2+1$ dimensional surface -- the defect -- while allowing
the photons to explore $d = 3+1$ dimensional space.   While Schwinger solved $1+1$ 
dimensional massless QED over fifty years ago \cite{Schwinger:1962tp}, these same techniques can be used to solve
a modified version where the electrons are constrained to $p=1+1$ dimensions but the photon is free
to propagate in more \cite{Fraser-Taliente:2024lea}.  

\vskip 0.1in

Just as conformal field theories play a special role as landmarks in the renormalization group space of QFTs,
defect conformal field theories play a similar role for defect QFTs.  They provide a richer set of endpoints  
for distinct  bulk and defect  renormalization group flows.  
The question we explore in this work is what types of defect conformal field theories exist when the bulk
is constrained to be a free field theory.  In the two examples mentioned above, the bulk theory is free 
electricity and magnetism, 
while here we shall for the most part be modest in our aims, focusing on the case of a massless free scalar
theory in the bulk.

\vskip 0.1in

To fix notation and definitions, we assume the defect lives on ${\mathbb R}^{1,p-1}$ Minkowski space which
is embedded in a ${\mathbb R}^{1,d-1}$ dimensional spacetime (in a mostly plus notation for the metric).  
The presence of the defect must break
some of the conformal $SO(d,2)$ symmetry of the bulk.  By a conformal defect, we mean that there is a 
$SO(p,2) \times SO(d-p)$ symmetry
where $SO(p,2)$ is the residual conformal symmetry preserved by the defect while
$SO(d-p)$ is the part of the bulk rotational group that leaves the defect in a fixed position.  
We also define the codimension $q \equiv d-p$.  

\vskip 0.1in

Ref.\ \cite{Lauria:2020emq} initiated a classification program for defects in bulk free theories.
In the case where the bulk contains only a real scalar field $\phi$, 
they make two inter-related observations and prove a triviality condition for operators in the defect expansion of $\phi$.
Recall the defect operator expansion (DOE) expresses $\phi$ as a sort of Taylor series 
sum over defect primary operators and their descendants.
Observation one is that the equation of motion for $\phi$ severely restricts the spectrum of operators in the DOE, providing a linear relation 
between their spin and scaling dimension.  Observation two concerns three point functions of $\phi$ and two defect operators.
The equation of motion and regularity (around unphysical singularities) in general imply a linear relation between the scaling dimensions of the two defect operators
and the scaling dimension of an operator in the DOE of $\phi$, sometimes called a ``double twist condition''.  
The triviality condition that follows from these assumptions is that there is a sector of defect operators, including the operators in the
DOE of $\phi$, which are all generalized free fields (GFF), i.e.\ their correlation functions follow from Wick's Theorem.
One of us demonstrated that the above conclusions generalize straightforwardly also to surface defects in Maxwell theory
\cite{Herzog:2022jqv}, suggesting a general pattern that may persist for conformal defects in free field theories generally.

\vskip 0.1in

There are however important exceptions to these statements.  
As discussed in ref.\ \cite{Lauria:2020emq}, 
for low lying spins occasionally unitarity permits a second operator in
the defect expansion of $\phi$ with a different scaling dimension.  In this case, the double twist condition that follows from a study
of three point functions no longer holds.  The proof of existence of a GFF sector then also fails.  These extra operators always
appear in the case of boundaries and interfaces (with $q=1$), which in turn 
allow for a much richer set of boundary conformal field theories, both for free scalars \cite{Behan:2020nsf,Behan:2021tcn} and
a Maxwell field \cite{DiPietro:2019hqe}. 
 Further exceptions are carved out when $q=3$ and the spacetime dimension $d\geq 5$ and also for monodromy defects with 
$q=2$ and $d \geq 4$.  

\vskip 0.1in

Given the rich set of boundary conformal field theories involving free fields \cite{Behan:2020nsf,Behan:2021tcn,DiPietro:2019hqe}, 
it makes sense to examine the assumptions of 
refs.\ \cite{Lauria:2020emq,Herzog:2022jqv} closely.
For the particular case of line defects $p=1$, ref.\ \cite{Lauria:2020emq} made the assumption that the 
correlation functions were invariant under  time reversal
symmetry, or equivalently an inversion symmetry.  Revisiting this assumption is the purpose of this work.\footnote{%
Note this assumption is much stronger than assuming the path integral (or theory) is invariant under time reversal.  Even if the theory is invariant, correlation functions of operators that transform nontrivially under time reversal in general will also transform nontrivially, in a way governed by Ward identities.
}   
(They also assume that the only dimension zero operator is the identity.)

\vskip 0.1in

Relaxing time reversal symmetry, we find that observation two -- the double twist condition -- fails.  
Specific to line defects, there is a cross ratio $\nu$ for bulk-defect-defect three point functions which is not invariant under time reversal but is invariant under the conformal group.
Expressed in terms of $\nu$, the bulk-defect-defect three point functions are smooth and no double twist condition need ever be imposed.
As a result, the proof \cite{Lauria:2020emq} of the GFF condition cannot be applied.  
 
 \vskip 0.1in
 
 We then design a 
line defect theory involving a massless, $3+1$ dimensional scalar, a massless one dimensional
complex fermion and a Yukawa type interaction between the two.  The theory  has operators -- the fermions -- and consequently correlation functions as well which transform nontrivially under time reversal.
It additionally has a dimension zero operator -- a charge $\bar \psi \psi$ -- 
which is not the identity.
Perhaps not surprisingly, it has a bulk-defect-defect three point function of the type expressly forbidden by the assumptions of \cite{Lauria:2020emq}.  
Curiously though this theory still has a GFF sector -- all the operators in the defect expansion of $\phi$ are GFF's, and their correlation functions with each other follow from Wick's Theorem.

\vskip 0.1in

The work is organized as follows. In section \ref{sec:correlations}, we adopt the  free bulk scalar field equation of motion as an input to constrain the bulk-defect two point function and the bulk-defect-defect three point function. The conformal blocks of the  three point function are assumed to depend on a cross ratio that is not invariant under inversion, and as a result it turns out 
there are no restrictions on the defect operator spectrum. In section \ref{sec:toymodel}, we develop a toy model with a fermionic line defect coupled to the bulk through a localized Yukawa term. 
The toy model has defect operators with nontrivial anomalous dimensions and allows for bulk-defect-defect three point functions that would have been forbidden by inversion symmetry.  The model remains very simple, however, in the sense that it contains a GFF sector and 
a field redefinition allows for an exact computation of all  correlation functions. We consider a few correlators of low transverse spin and find a precise match with the general result of section \ref{sec:correlations}. 
Section \ref{sec:discuss}  concludes with a discussion
about the possibility of an  extension to the Maxwell case.  Technical details and supplementary material are relegated to the appendices. Appendix \ref{app:nu} gives a more detailed discussion on the inversion broken cross ratio; appendix \ref{app:polarization} shows the steps of the construction of the polynomial representations of $SO(d-1)$ important for the three point function; appendix \ref{app:bulkdefectdefect}
computes the bulk-defect-defect correlation using an OPE approach; while appendix \ref{app:beta}
provides a two-loop calculations of the beta function of the Yukawa coupling and fermion wavefunction renormalization for the toy model.

\section{Correlation Functions from the Equation of Motion}
\label{sec:correlations}

The triviality of ref.\ \cite{Lauria:2020emq} is a statement about the defect operators in the DOE of the bulk scalar field $\phi$.
The claim is that any correlation function of these defect operators must follow from Wick's Theorem.  The fact that the correlation functions can be calculated in 
such a simple manner suggests that the field theory is that of generalized free fields, and is in that sense trivial.  In rough outline, their proof 
is to use a bulk-defect-defect 
three point function to establish a  restriction on the defect operator spectrum, and then to show this restriction, via a contour integral argument, implies triviality.
To examine a case excluded by their assumptions, let us then first take a closer look at bulk-defect-defect three point functions for line defects.

\vskip 0.1in

Conformal invariance fixes the form of these three point correlators up to a set of functions that depend on a single cross ratio, which is often conventionally written as
\begin{equation}
\label{eq: definition of zeta cross ratio}
\zeta \equiv \frac{(x_{12})^2 (x_{13})^2}{ |x_{23}|^2 y_1^2}
\,,
\end{equation}
where the bulk operator is inserted at $x_1 = (t_1, y_1)$ and the defect operators are inserted at $x_2= (t_2, 0)$ and $x_3 = (t_3, 0)$.
This $\zeta$ is invariant under the full $SO(p,2) \times SO(d-p)$ symmetry group, and is in fact invariant under time reversal $t \to -t$ as well, which is not part of $SO(p,2) \times SO(d-p)$.

\vskip 0.1in

Remarkably, for line defects $p=1$, $\sqrt{\zeta + 1}$ can be written in a way that does not involve any branch cuts for the time coordinates:

\begin{equation}
\label{eq:nucrossratio}
\nu \equiv  \frac{(t_2 - t_1)(t_1-t_3) + y_1^2}{(t_2 - t_3) |y_1|} \ .
\end{equation}
This new cross ratio $\nu$ is invariant under $SO(1,2)^+ \times SO(d-1)$ but picks up a sign under inversion (or equivalently under $t \to -t$).  In other words,
it is useful for writing three point functions for theories without inversion symmetry.  

\vskip 0.1in

Another way of seeing how this new cross ratio arises for line defects but not more generally is to use embedding space (or null cone formalism, see for example \cite{Rychkov:2016iqz}).
In this picture, a $p$-dimensional defect is lifted to a $p+2$ dimensional object where the conformal group acts linearly.
There is an epsilon tensor $\epsilon_{i_1 \cdots i_{p+2}}$  with $p+2$ 
indices which can be used for constructing invariant objects.  In the line defect case, we can saturate the indices with the location of the 
one bulk and two
defect operators.  The parity operation in the embedding space is inversion.
That $\epsilon$  
picks up a sign under parity pushes down to the fact that $\nu$ picks up
a sign under inversion.  That much of what we discuss here is specific to line defects is related then to the facts that this epsilon tensor
has $p+2$ indices, and we are interested specifically in constraints on three point functions.
(These types of cross ratios are discussed briefly in
refs.\ \cite{Billo:2016cpy, Costa:2011mg}.)

\vskip 0.1in

We try to work for the most part in Lorentzian signature, but there is a related Euclidean cross ratio $\nu = i \nu_E$ which we can write out as
(with the usual identification $t = -i \tau$)
\begin{equation}
\label{eq: definition of euclidean nu cross ratio}
\nu_E = \frac{(\tau_1 - \tau_2)(\tau_1-\tau_3) + y_1^2}{(\tau_2 - \tau_3) |y_1|} \ .
\end{equation}
For both cross ratios, their behavior is singular when the two defect operators approach each other $\tau_2 \to \tau_3$ or when the bulk operator approaches the line $|y_1| \to 0$.  
In the Lorentzian case, we also expect singular behavior when $x_1$ is on the forward or backward lightcones of the defect insertions $x_2$ and $x_3$: $t_1 = \mp |y_1| + t_2$ and $t_1 = \pm |y_1| +t_3$.  In these limits $\nu \to \pm 1$, and we will see in the correlation functions we compute below corresponding singular behavior at these special values of $\nu$.  A more complete discussion of this cross ratio is included as appendix \ref{app:nu}.

\subsection{Two Point Functions and the DOE}

Before tackling the three-point function, we review the simpler case of two-point functions.
We constrain the form of $\langle \phi(x_1) \hat O_{\ell}(t_2) \rangle$ and 
$\langle \phi(x) \phi(x')\rangle$
correlation functions where $\phi(x)$ is a free scalar that satisfies the Klein-Gordon equation $\Box \phi = 0$ and 
$ \hat O_{\ell}(t)$ is a defect operator in a traceless symmetric representation of $SO(d-1)$.   
The notation $\hat O_{\ell}(t)$ is shorthand for $p^{i_1} \cdots p^{i_\ell} O_{i_1 \cdots i_\ell}(t)$ where $p$ is assumed to be a light-like vector.  This polarization vector associated with $\hat O_\ell(t)$ helps in handling the tracelessness condition on the representation in a simple way.
In addition to the light-like vector $p_a$ for each defect insertion, we also need a normal vector $n = y_1 / |y_1|$.  Let $\Delta = \frac{d-2}{2}$ be the scaling dimension of $\phi$ and $\Delta_\ell$ the scaling dimension of $ \hat O_{\ell}(t_i)$. Interactions on the defect can never renormalize $\Delta$ away from its free field value. 

\vskip 0.1in

Starting with the two point function, we find
\begin{equation}
\label{eq:bulkdefect}
\langle \phi(x_1) \hat O_{\ell}(t_2) \rangle = \mB_{\phi \hat O_\ell} \frac{(n\cdot p_2)^\ell}{|x_1-x_2|^{2 \Delta_\ell} |y_1|^{\Delta - \Delta_\ell}}
\end{equation}
where $\mB_{\phi \hat O_\ell}$ is a normalization constant.
This object satisfies the free field equation of motion $\Box \phi = 0$ only if $\Delta_\ell = 1 - \Delta-\ell$ or $\Delta_\ell = \Delta+\ell$.  In $d=4$, we have the specific cases $\Delta_\ell = -\ell$ and $\Delta_\ell = \ell + 1$.  Unitarity $\Delta_\ell \geq 0$ restricts the $\Delta_\ell =-\ell$ solutions to the case $\ell=0$.  

\vskip 0.1in

This result for the bulk-defect two point function in turn restricts the type of defect operators that can show up in the DOE of the bulk field $\phi$.  In $d=4$, we expect in general the existence of a scalar operator with $\ell=0$ and dimension $\Delta_\ell = 0$ and then a tower of transverse spinning operators with $\ell = 0, 1, 2, \ldots$ and $\Delta_\ell = 1 + \ell$.  Given the right assumptions, it is this tower of operators, of the form $\der_\perp^\ell \phi$, which are expected to be GFF's. 

\vskip 0.1in

In fact, we can use the bulk-defect two-point function to derive
the defect OPE for $\phi(x)$.  (The decomposition is similar to the boundary OPE in ref.\ \cite{McAvity:1995zd}.)
In detail
\begin{align}\label{eq:phidOPE}
\phi (t, y) &= \sum_{\hat O_\ell} \frac{
\mB_{\phi \hat O_\ell}}{\mN_{\hat O_\ell \hat O_\ell}} \frac{1}{|y|^{\Delta - \Delta_{\ell}}} 
\mD^{\Delta_{\ell}}(|y|^2 \der_t^2)\hat O^\ell(t)\,,\\ 
\mD^{\Delta_{\ell}}(|y|^2 \der_t^2)& = \sum_m \frac{1}{m!\, (\Delta_{\ell} +1/2)_m} \left( \frac{1}{4} |y|^2 \der_t^2\right)^m\,,
\end{align}
with $\mN_{\hat O_l \hat O_l}$ being the two-point function normalization for
the defect field $\hat O_\ell(t)$, and $\mB_{\phi \hat O_\ell}$ the bulk-defect two-point function coefficient. 
We have introduced a subtle shorthand here:
while $\hat O_\ell(t) \equiv p^{i_1} \cdots p^{i_\ell} O_{i_1 \cdots i_\ell}(t)$, instead with the superscript
 $\hat O^\ell(t) \equiv n^{i_1} \cdots n^{i_\ell} O_{i_1 \cdots i_\ell}(t)$.
As a quick confirmation of the defect OPE, we use it to evaluate the bulk-defect two-point function
\begin{align}
\langle \phi (x_1) \hat O_{\ell} (t_2) \rangle
&= \sum_{\hat O_{\ell'}} \frac{
\mB_{\phi \hat O{_\ell'}}}{|y_1|^{\Delta - \Delta_{\ell'}}} \mD^{\Delta_{\ell'}} (|y_1|^2 \der_{t_1}^2)  \frac{\delta_{\hat O_\ell \hat O_{\ell'}} (n 
\cdot p_2)^{\ell}}{|t_{12}|^{2 \Delta_{\ell}}}\no 
\ , 
\end{align}
which after carrying out the sum gives the expected form (\ref{eq:bulkdefect}) 
fixed by conformal symmetry. 

\vskip 0.1in

Although we will not use it in what follows, for completeness we can use the defect OPE to decompose the bulk-bulk two point function\footnote{%
 (B.8) of \cite{Dolan:2000ut} is useful for evaluating the resulting double sum.
}
\begin{align}
\langle \phi(x_1) \phi(x_2) \rangle = 
\sum_{\hat O_\ell} \frac{(
\mB_{\phi \hat O_\ell})^2
/\mN_{\hat O_\ell \hat O_\ell}
}{(|y_1||y_2|)^{\Delta - \Delta_{\ell}}} \,_2F_1 \left(\frac{\Delta_{\ell}}{2},\frac{\Delta_{\ell}+1}{2},\frac{\Delta_{\ell}+2}{2}; \frac{1}{\xi^2} \right) \frac{\bar C^{\left(\frac{d-3}{2}\right)}_\ell(n_1\cdot n_2)}{( -t_{12}^2 + y_1^2 + y_2^2)^{\Delta_{\ell}}}\,,
\end{align}
where $t_{12} = t_1 -t_2$, and $\bar C_{\ell}^{(\alpha)} =   C_\ell^{(\alpha)}(x)/\frac{2^\ell (\alpha)_\ell}{\ell!}$ are rescaled from the  Gegenbauer polynomials (sometimes called ultraspherical polynomials).  These polynomials have the nice property that they are eigenfunctions of the $SO(d-1)$ Casimir operator acting on either $n_1$ or $n_2$ with eigenvalue $\ell (\ell+d-3)$.
Also, we have introduced the cross ratio $\xi = (-t_{12}^2 + y_1^2 + y_2^2)/(2 |y_1| |y_2|)$.
The bulk two-point function in a defect theory typically depends on two cross ratios.  The other cross ratio in this case is $(n_1 \cdot n_2)$.

\subsection{The Bulk-Defect-Defect Three Point Function}

Next we look at the bulk-defect-defect three point function $\langle \phi(x_1) \hat O_{\ell_2} (t_2) \hat O_{\ell_3}(t_3) \rangle$.
This object should decompose into a sum over conformal blocks, where each block corresponds to exchange of a primary operator $\hat O_{\ell}$ (along with its descendants)
that is both in the DOE of the bulk field $\phi$ and the  OPE of the two defect fields $\hat O_{\ell_2}$ and $\hat O_{\ell_3}$.
Given the $SO(2,1) \times SO(d-1)$ symmetry, we expect these conformal blocks to be eigenfunctions of the Casimir operator of this group.  The eigenvalues will have a contribution $\Delta_{\ell}(\Delta_{\ell}-1)$ from the $SO(2,1)$ part and a contribution $\ell(\ell+d-3)$ from the $SO(d-1)$ part.  

\vskip 0.1in

Given the discussion about two point functions above, for a free bulk field $\phi$ there is an immediate further restriction on $\Delta_\ell$.  All of the defect fields in the DOE of $\phi(x)$ obey the relation $\Delta_{\ell} = \Delta+\ell$ or $1-\Delta-\ell$.  In the special case $d=4$, for a unitary theory 
the sum over conformal blocks restricts to $\Delta_\ell = \ell+1$, and in the special case $\ell=0$, also $\Delta_0 = 0$.   

\vskip 0.1in

There is also a restriction on the allowed values of $\ell$. The OPE of $\hat O_{\ell_2}$ and $\hat O_{\ell_3}$  should produce operators in the tensor product representation
of $\ell_2$ and $\ell_3$.  
Decomposing this tensor product into irreducibles  produces all symmetric traceless representations with a number of indices which varies from $|\ell_2 - \ell_3|$ to $\ell_2 + \ell_3$.  Thus if $\ell_2 \geq \ell_3$, $\ell = \ell_2 + \ell_3 - m$ where $m=0, 1, \ldots, 2 \ell_3$.
In fact, because the indices contract pairwise to produce new representations, we expect $m$ to be even.  In the special case of $SO(3)$, however, there is an additional epsilon tensor $\epsilon_{ijk}$ we can use to obtain odd $m$ as well.

\vskip 0.1in

Generically then for a free field $\phi(x)$, this bulk-defect-defect three point function can be decomposed into a sum of conformal blocks corresponding to each allowed value of $\ell$ and the corresponding allowed values of $\Delta_\ell$.  The sum over even $m$ described above has the form\footnote{%
Interpreted in Euclidean signature, this expression is unambiguous.
In Lorentzian signature, morally the expression is the time ordered Feynman Green's function obtained by analytically continuing from Euclidean signature, but we have suppressed factors of $i \epsilon$.
}
\begin{equation}
\label{eq:predictedthreepoint}
\langle \phi(x_1) \hat O_{\ell_2} (t_2) \hat O_{\ell_3}(t_3) \rangle = \frac{(n\cdot p_2)^{|\ell_2 - \ell_3|} (p_2 \cdot p_3)^{\ell_3}}{|x_{12}|^{2 \Delta_2} |x_{13}|^{2 \Delta_3}
|y_1|^{\Delta - \Delta_2 - \Delta_3}} \sum_{\Delta_\ell} c_{\Delta_\ell}  f_{\Delta_\ell}(\nu) h_\ell(\chi)  \ ,
\end{equation}
depending on the cross ratios 
\begin{equation}
\chi \equiv \frac{(n \cdot p_2) (n \cdot p_3)}{p_2 \cdot p_3} \ ,
\end{equation}
and $\nu$ we saw before.  
The $c_{\Delta_\ell} = {\mathcal B}_{\phi \hat O_\ell} {\mathcal C}_{\hat O_{\ell_2} \hat O_{\ell_3} \hat O_\ell} / {\mathcal N}_{\hat O_\ell \hat O_\ell}$ can be decomposed using the OPEs as a rational expression of two and three point function coefficients.
In $d=4$, we may add also a second sum corresponding to odd values of $m$:
\begin{equation}
\frac{(n\cdot p_2)^{|\ell_2 - \ell_3|} (p_2 \cdot p_3)^{\ell_3-1} \epsilon_{ijk} n^i\, p_2^j\, p_3^k }{|x_{12}|^{2 \Delta_2} |x_{13}|^{2 \Delta_3}
|y_1|^{\Delta - \Delta_2 - \Delta_3}} \sum_{\Delta_\ell} \tilde c_{\Delta_\ell}  f_{\Delta_\ell}(\nu) \tilde h_\ell(\chi)  \ .
\end{equation}

\vskip 0.1in

To check this construction and find explicit forms for $f_{\Delta_\ell}(\nu)$ and $h_\ell(\chi)$,
we look for solutions of
\begin{equation}
\Box_{x_1} \left(\frac{(n\cdot p_2)^{|\ell_2 - \ell_3|} (p_2 \cdot p_3)^{\ell_3}}{|x_{12}|^{2 \Delta_2} |x_{13}|^{2 \Delta_3}
|y_1|^{\Delta - \Delta_2 - \Delta_3}} F(\nu, \chi)\right) = 0 \ , 
\end{equation}
i.e.\ we are checking the condition that $\phi(x)$ really is a free field.
(A separate derivation of this bulk-defect-defect three point function using the OPE is in appendix \ref{app:bulkdefectdefect}.)
The resulting second order partial differential equation can be solved by separation of variables
$F(\nu, \chi) = f(\nu) h(\chi)$, yielding
two ordinary second order differential equations.
The ``Casimir equation" for $f(\nu)$ is
\begin{equation}
\label{eq:2ndorder}
(1-\nu^2) f''(\nu) + 2(\Delta_+-1) \nu f'(\nu) + \left(  \Delta_+  + \frac{\Delta_+^2 \nu^2 - \Delta_-^2}{1-\nu^2}
\right) f(\nu) =  -\Delta_\ell(\Delta_\ell-1) f (\nu) \ , 
\end{equation}
where $\Delta_\pm = \Delta_2 \pm \Delta_3$ and $\Delta_\ell = \ell+\Delta$.
The ``Casimir equation" for $h(\chi)$, on the other hand, is
\begin{align}
\label{eq:Cash}
2(1-2 \chi) \chi h''(\chi) 
+  2 \left(1 + (\ell_2 - \ell_3)(1-2 \chi) +  \chi(1-d) \right) & h'(\chi) + \\
- (\ell_2 - \ell_3)(\ell_2-\ell_3 + d-3) h(\chi)
&=  - \ell(\ell+d-3) h(\chi) \ . \nonumber
\end{align}

The solutions of both differential equations are easily although not particularly informatively expressed as hypergeometric functions.  For example, for $h(\chi)$, we immediately find the polynomial solutions
\begin{equation}
\label{eq:hchi}
h(\chi) = {}_2 F_1 
\left(
\frac{-\ell+\ell_2 - \ell_3}{2}, 
\frac{d-3}{2} + \frac{\ell + \ell_2 - \ell_3}{2}, 1 + \ell_2 - \ell_3; 2 \chi 
\right) \ .
\end{equation}
These particular hypergeometric polynomials are called Jacobi polynomials
\begin{equation}
{}_2 F_1
\left( a + \frac{d-3}{2}, -b, a-b+1; 2 \chi \right)
= \frac{b!}{(a-b+1)_b} P_b^{(a-b, \frac{d-5}{2})}(1-4 \chi) \ .
\end{equation}
In the particular case $d=4$, we find the Legendre polynomials:
\[
P_{b}^{\bigl(a-b,-\frac12\bigr)}\bigl(1-4x\bigr)
=
(-1)^{\,a-b}
\frac{a!\,(2b)!}{b!\,(2a)!}
2^{\frac{a-b}{2}}
x^{\frac{b-a}{2}}
P_{\,a+b}^{\,a-b}\!\bigl(\sqrt{1-2x}\bigr) \ ,
\]
which transform nicely under $SO(3)$.
In appendix \ref{app:polarization}, we provide an alternate construction of the $h_\ell(\chi)$ and $\tilde h_\ell(\chi)$ that makes their transformation properties under $SO(d-1)$ transparent.  

\vskip 0.1in

The differential equation for $f(\nu)$ 
has solutions
\begin{equation}
\label{eq:Legsols}
f(\nu) = ( \nu^2-1)^{\frac{\Delta_2 + \Delta_3}{2}} \left( c_1 P_{\Delta_\ell-1}^{\Delta_2 - \Delta_3}( \nu ) 
+ c_2 Q_{\Delta_\ell-1}^{\Delta_2 - \Delta_3} (\nu)  \right) \ .
\end{equation}
(Note this form works for both even and odd $\ell$.)
The existence of two solutions corresponds to the fact that both the conformal block with $\Delta_\ell$ and its shadow dual with dimension $1-\Delta_\ell$ satisfy the same Casimir equation.  In the special case $d=4$,
we have $\Delta_\ell = \ell+1$ and $-\ell$.  
Except in the case 
$\ell=0$, we can discard the solution corresponding to 
$\Delta_\ell = - \ell$ because it is below the unitarity bound. 
The selection of what amounts to particular boundary conditions is more easily carried out in the $\zeta =  \nu^2-1$ variable, as 
was done originally in ref.\ \cite{Lauria:2020emq}
in the context of $p$-dimensional defects in a $d$-dimensional bulk.
This transformed differential equation has the 
hypergeometric solutions
\begin{eqnarray}
f(\nu) &=& c_1' \zeta^{(\Delta_+ + \Delta_\ell-1)/2} 
{}_2 F_1 
\left(
-\frac{\Delta_- + \Delta_\ell-1}{2}, \frac{\Delta_- - \Delta_\ell+1}{2}, \frac{3}{2} - \Delta_\ell; - \frac{1}{\zeta}
\right) \nonumber \\
&&
+ c_2' \zeta^{(\Delta_+ - \Delta_\ell)/2}
{}_2 F_1 
\left(
\frac{-\Delta_- + \Delta_\ell}{2}, \frac{\Delta_- + \Delta_\ell}{2}, \frac{1}{2} + \Delta_\ell; - \frac{1}{\zeta}
\right)
\end{eqnarray}
which are of course equivalent to the Legendre functions above, as they satisfy the same differential equation after a change of variable.
If we analyze the bulk-defect-defect three point function in the coincident limit $t_2 \to t_3$, we find the scaling of the schematic form
\[
c_1'  t_{23}^{-\Delta_+ - \Delta_\ell+1} + c_2' t_{23}^{-\Delta_+ + \Delta_\ell  } \ .
\]
In this limit, we expect to pick out a $\hat O_{\ell}(t)$ in the operator product expansion of the two $\hat O_{\ell_2}$ and $\hat O_{\ell_3}$ operators which should lead to the scaling corresponding to the $c_2'$ behavior above, setting $c_1'=0$.  
If we analyze the correlation function instead in the limit $y_1 \to 0$, we find
\[
c_1' y_1^{-\Delta -\Delta_\ell +1} + c_2' y_1^{-\Delta + \Delta_\ell}\ .
\]
Here we expect to pick out a $\hat O_{\ell}(t)$ in the defect operator expansion of $\phi$, again picking out the $c_2'$ solution. 

\vskip 0.1in

An issue pointed out by ref.\ \cite{Lauria:2020emq} 
that was critical in their proof 
is that the hypergeometric multiplying $c_2'$ will in general have a square root singularity
$\sqrt{\zeta + 1}$ near $\zeta=-1$ unless the spectrum is restricted.  If we impose the ``double twist condition''
$\Delta_2 = \Delta_\ell + \Delta_3 + 2n$
or $\Delta_3 = \Delta_\ell + \Delta_2 + 2n$ where $n = 0, 1, 2, \ldots$, then the hypergeometric becomes a polynomial in $\zeta$ and there can be
no $\sqrt{\zeta+1}$ singularity.  However, we know 
in the $p=1$ case we can make the replacement $\sqrt{\zeta + 1} 
\to \nu$.  Thus there is no true square root singularity here.  Indeed,
the Legendre functions (\ref{eq:Legsols}) are perfectly well behaved at $\nu = 0$. 
Instead what happens is that if the spectrum is not
restricted, the correlation functions of the corresponding defect CFT will not be invariant under this inversion symmetry.

\section{Yukawa Interaction on the Line}
\label{sec:toymodel}

We introduce the following fermionic line defect coupled to a free scalar $\phi(x) = \phi(t, \vec x)$
\begin{equation}
\label{eq:themodel}
\int_{\vec x=0} dt  \left[ \bar \psi(i \partial_t + g \phi) \psi  + h \phi \right] - \frac{1}{2} \int d^d x \, (\partial_\mu \phi) (\partial^\mu \phi) \ .
\end{equation}
The real numbers $g$ and $h$ are coupling constants.  Although the presence of the fermions make the action look complicated, at core we have a degenerate two-state quantum mechanical system coupled to an external bath of scalar particles.
Our main interest is in the case $d = 4-\epsilon$, 
where we can expect the couplings $g$ and $h$ to be close to marginal when $\epsilon \ll 1$,
and the theory to be close to conformal.  In fact, we will see that the critical case is precisely when $\epsilon=0$ and that $g$ is exactly marginal.  
 To make the defect fermionic, we assume in the decoupled limit $g=h=0$ that the $\psi$-fields satisfy the equal time anti-commutation relations
$ \{ \psi(t),  \bar  \psi(t) \} = 1$.  Such a model was also considered in 
\cite{Allais:2014fqa,Bashmakov:2024suh}, while  similar bosonic models appeared recently in 
\cite{Cuomo:2022xgw,Komargodski:2025jbu} (see also \cite{Vojta1999}).  
The exact treatment of a similar system in ref.\ \cite{Anninos:2016klf} inspired our choice of this particular model.

\vskip 0.1in

Remarkably, this interacting field theory can be solved via the following nonlocal field redefinition,
\begin{equation}
\label{eq:trivialization}
\psi(t) = \exp \left( i g \int^t \phi(\tau) d\tau \right) \psi'(t)  \ , \; \; \; \bar \psi(t) = \exp \left(-i g \int^t \phi(\tau) d\tau \right) \bar \psi'(t),
\end{equation}
where the lower bound of the integral sets a scale.  The transformation 
yields the following decoupled system
\begin{equation}
\label{eq:decoupled}
\int_{\vec x=0} dt  \left[ i \bar \psi' \partial_t  \psi'  + h \phi \right] - \frac{1}{2} \int d^d x \, (\partial_\mu \phi) (\partial^\mu \phi) 
\end{equation}
of two free fields. Despite its relation to a free system, the original model (\ref{eq:themodel})
has many properties
characteristic of interacting quantum field theories, for example anomalous dimensions for the fermionic operators $\psi$ and $\bar \psi$.  At the same time, the map to a free system allows us to check and make sense of results we obtain through a perturbative analysis.

\vskip 0.1in

The equations of motion that follow from the original model (\ref{eq:themodel}) are
\begin{equation}
\Box \phi  + (g \bar \psi \psi  + h) \delta^{(d-1)}(x^i) = 0 \ , \; \; \; i \partial_t \psi + g \phi \psi = 0 \ ,
\label{eq:eom}
\end{equation}
indicating that $g \bar \psi \psi +h$ acts as a delta function source for the bulk field.  Assuming the vacuum on the defect line satisfies $\psi(0) | 0 \rangle = 0$, then the one point function of the composite operator $\langle \bar \psi \psi(t) \rangle = 0$ should vanish, and only $h$ will produce a bulk response:
\begin{equation}
\label{eq:phibg}
\langle \phi(t,y) \rangle = \frac{h}{(d-3) \Vol(S^{d-2})} \frac{1}{|y|^{d-3}}  \stackrel[d\to 4]{}{\longrightarrow} \frac{h}{4 \pi |y|}\ ,
\end{equation}
where $|y|$ is the distance from the defect line.  

\vskip 0.1in

To obtain the two point functions, it is useful to start from the decoupled system (\ref{eq:decoupled}).
Setting $h=0$, 
 we find the familiar Feynman Green's functions for the fermion and scalar:
\begin{eqnarray}
\langle T \phi(x) \phi(x') \rangle &=& \frac{1}{(d-2) \Vol(S^{d-1})} \frac{1}{((x-x')^2 )^{\frac{d-2}{2}}} 
 \stackrel[d\to 4]{}{\longrightarrow} \frac{1}{4 \pi^2} \frac{1}{(x-x')^2}\ , \\
\langle T \psi'(t) \bar \psi'(t') \rangle &=& \Theta(t-t') \ .
\end{eqnarray}
The theta function result for the fermion follows trivially from the definition of the Feynman Green's function, the anticommutation relation for $\psi'$ and $\bar \psi'$, and the definition of the vacuum $\psi'(0) | 0 \rangle = 0$.  

\vskip 0.1in

Mapping these free results back to the interacting model, we see the two-point function for $\phi$ is unchanged while $\psi$ picks up an anomalous dimension:
\begin{equation}
\left \langle \exp \left( i g \int^{t} \phi(\tau) d\tau \right)   \psi'(t) \exp \left(- i g \int^{t'} \phi(\tau) d\tau \right) 
\bar \psi'(t') \right \rangle \sim |t-t'|^\frac{-g^2}{4\pi^2} \Theta(t-t') \ .
\end{equation}
Indeed the scaling dimension of the fermion and its complex conjugate is no longer zero in the interacting model but instead equal to
\begin{equation}
\Delta_\psi = \Delta_{\bar \psi} = \frac{g^2}{8 \pi^2} \ .
\end{equation}
A simple way of thinking about the action of these fermions is that they create or annihilate the line defect associated with integrating $g \phi$ over the line.
Indeed, this value for the scaling dimension of $\psi$ and $\bar \psi$ agrees with the dimension of such defect changing operators computed in ref.\
\cite{Allais:2014fqa,Cuomo:2024psk}.

\vskip 0.1in

The scalar in contrast keeps its unperturbed scaling weight $\Delta_\phi = 1$.  Indeed, we do not expect a local defect to be able to change a bulk scaling dimension, for the tail to wag the dog.
Because the rescaling factors cancel out in $\bar \psi \psi$, the conformal dimension of $\bar \psi \psi$ is also unchanged, $\Delta_{\bar \psi \psi} = 0$.  Indeed as we will review shortly, $\bar \psi \psi$ has the interpretation of a current operator (or really just a charge in this one dimensional case), whose scaling dimension should be protected against the effects of interaction.

\subsection{Symmetries}

For $h=0$, there is a symmetry of (\ref{eq:themodel}) associated with the transformations
\begin{align}
\begin{split}
& \psi \to e^{i g \alpha} \psi \ , \; \; \; \bar \psi \to e^{-i g \alpha} \bar \psi \ , \\
& \phi \to \phi + \partial_t \alpha \ .
\label{eq:sym}
\end{split}
\end{align}
The fermionic part of the action is invariant under this transformation for all $\alpha(t)$, while the kinetic
term for the scalar requires that  $\partial_\mu \partial_t \alpha$ is zero; in other words
\begin{equation}
\alpha = t \alpha_1 + \alpha_0  \ .
\end{equation}
Note a contribution of the form $t x^\mu \beta_\mu$ to $\alpha$ leads to a shift of the Lagrangian by an overall constant, and is thus not  a symmetry.\footnote{%
The constant here is order $\beta^2$.
 Infinitesimally, the action shifts by a total derivative, but to maintain this shift as a finite symmetry requires more exotic conditions on $\beta$ that we will not consider.
} 
We could make $\alpha_0$ a function of the spatial $x^i$, but this extra information is redundant and not incorporated in the symmetry transformation rules.
Turning on $h$, the symmetry associated with $\alpha_1$ 
is in general broken.\footnote{%
  If we make the time circle compact however with circumference $\beta$, then we can partially restore this symmetry.
First, on physical grounds we should make sure that rotating the fermions by angle $\alpha(0)$ is equivalent to rotating them by angle $\alpha(\beta)$, up to a sign $\pm 1$, 
$g \alpha_1 \beta \in \pi {\mathbb Z}$.
We further require for this $h \phi$ source term not to affect the path integral that $h \beta \alpha_1 \in 2 \pi {\mathbb Z}$.  Thus 
$g/h$ must be rational to preserve any remnant of the $\alpha_1$ shift symmetry, and the allowed values of $\alpha_1$ are restricted.
}

\vskip 0.1in

There are currents associated with these symmetries.  Associated with $\alpha_0$, we have the defect charge $j^0 = \bar \psi \psi$, which is conserved $\partial_0 j^0 = 0$.
Associated with the bulk shift symmetry $\alpha_1$, we have the bulk current
\begin{equation}
J^\mu = \partial^\mu \phi \ , 
\end{equation}
which is conserved away from the defect.  Including the defect, we find instead
\begin{equation}
\partial_\mu J^\mu =  - (g \bar \psi \psi + h) \delta^{d-1}(x^i) \ ,
\end{equation}
where $T =- (g \bar \psi \psi + h)$ is sometimes called a tilt operator.  In this particular case, the tilt operator is also conserved, $\partial_t T = 0$, which in turn implies $\partial_0 \partial_\mu J^\mu = 0$.

\vskip 0.1in

Consider for the moment a $d$ dimensional region $M$ inside our space-time with $d-1$ dimensional boundary $\partial M$, setting also $h=0$ for simplicity.  We can integrate the Hodge dual of our current over the boundary of the region, $\int_{\partial M} \star J$.  By Stokes' Theorem, this integral is equal to $\int_M d \star J$.  From our construction, the divergence $d \star J$ is zero everywhere except on the line defect, where it evaluates to the conserved charge $g \bar \psi \psi = g q$. The integral over $M$ thus reduces to an integral over the line defect inside $M$, and the result is the charge times the length of the line defect $L$ contained in $M$, which we can write as $L\, g q$.\footnote{%
The structure here is similar to the primary and secondary currents discussed in ref.\ \cite{Hull:2023iny}.  In that set-up, there is a secondary anti-symmetric two-form current $J_{\mu\nu}$ whose divergence yields a primary, conserved one-form current $j_\mu$, i.e.\ $\partial^\mu J_{\mu\nu} = j_\mu$ where $\partial^\mu j_\mu = 0$.  
}

\vskip 0.1in

There is a Ward identity associated with the $\alpha_0$ shift symmetry.  Consider the correlation function
\begin{equation}
\langle j^0(t_1) \psi(t_2) \bar \psi(t_3) \rangle \sim \frac{
\Theta(t_2 - t_1) \Theta(t_1-t_3)}
{|t_2-t_3|^{\frac{g^2}{4 \pi^2}}} \ ,
\end{equation}
where we have evaluated the result using the nonlocal field redefinition.
The conventional Ward identity is obeyed:
\begin{equation}
\partial_t \langle j^0(t) \psi(t_2) \bar \psi(t_3) \rangle =
-\delta(t_2-t) \langle \psi(t_2) \bar \psi(t_3) \rangle
+ 
\delta(t-t_3) \langle \psi(t_2) \bar \psi(t_3) \rangle \ .
\end{equation}

\vskip 0.1in

This Ward identity in QED is often used to argue that the wave function renormalization of the coupling and the fermions are related, and hence that the beta function is determined solely by the wave function renormalization of the photon.  Similarly, in our case, the beta function for the coupling can only come from the scalar wave function renormalization.  But here, the scalar is a bulk field which cannot be renormalized by the defect.  In fact there is a more pedestrian way of seeing that the bulk scalar is not renormalized.  Any fermion loop in this theory must vanish because of the cyclic product of theta functions.  Thus the tree level scalar propagator in this theory is not corrected at all. 
In other words, this example is an honest line defect CFT in $d=4$ with an exactly marginal parameter $g$.
In fact, $h$ does not affect the $\alpha_0$ shift symmetry, and 
should be a second marginal parameter whose
only effect is to give $\phi$ a bulk expectation value.

\vskip 0.1in

Before we realized the nonlinear field redefinition, we worked out the beta function in the $h=0$
theory to two loops.
We include this calculation in appendix \ref{app:beta}.  It is instructive and reassuring to see how the Feynman diagram calculation reproduces in a highly nontrivial way the results described above, especially since in many theories one does not have an exact solution to compare to.

\vskip 0.1in

The existence of a tilt operator is usually associated with the existence of a conformal manifold \cite{Drukker:2022pxk,Herzog:2023dop}.  Here the situation is more subtle because the current $J^\mu$ does not have the canonical dimension three of a conserved current in a 4d conformal field theory.  The symmetry transformation $\phi \to \phi + \alpha_1$ involves a scale dependent quantity $\alpha_1$ and thus does not commute with the conformal group.   

\vskip 0.1in

To end this subsection, we make some remarks about discrete symmetries.
There exists an anti-unitary time reversal transformation which leaves the fermion kinetic term untouched:
\begin{equation}
    {\mathcal T} i {\mathcal T}^{-1} \to -i \ , \; \; \;
    {\mathcal T} \psi {\mathcal T}^{-1} \to \bar \psi \ , \; \; \;
    {\mathcal T} \bar \psi {\mathcal T}^{-1} \to \psi \ .
\end{equation}
The interaction term, however, does not transform well: $\bar \psi \psi \to \psi \bar \psi = 1- \bar \psi \psi$.  
One way out is to set $h=1/2$ (half filling) and try to insist that the field $\phi$ is a pseudoscalar, picking up a sign both under time reversal and parity.  Time reversal symmetry is recovered at the price of breaking parity symmetry.

\subsection{Three Point Functions}

We first compute $\langle \phi(x_1) \psi(t_2) \bar \psi(t_3) \rangle$ exactly using the 
nonlocal mapping:
\begin{equation}
\langle \phi(x_1) \psi(t_2) \bar \psi(t_3) \rangle
= {\mathcal C} \frac{ig}{4\pi^2} \frac{
-\tanh^{-1} \frac{|y_1|}{t_1 - t_2} +
\tanh^{-1} \frac{|y_1|} {t_1 - t_3}
}
{|t_2-t_3|^{\frac{g^2}{4 \pi^2}} |y_1|}
\Theta(t_2-t_3)
\end{equation}
which simplifies via a tangent addition formula to
\begin{equation}
 {\mathcal C} \frac{ig}{4\pi^2}  \frac{
\tanh^{-1} \left(\frac{1}{\nu} \right)
}
{|t_2-t_3|^{\frac{g^2}{4 \pi^2}} |y_1|}
\Theta(t_2-t_3)
=
 {\mathcal C}\frac{ig}{4\pi^2}  \frac{(\nu^2 -1)^{\frac{g^2}{8\pi^2}} 
\tanh^{-1} \left(\frac{1}{\nu} 
\right)
}
{|x_{12}|^{\frac{g^2}{4\pi^2}}
|x_{13}|^{\frac{g^2}{4\pi^2}} 
|y_1|^{\Delta - \frac{g^2}{4\pi^2}}}
\Theta(t_2-t_3) \ .
\label{eq:3ptfinal}
\end{equation}
We have included a normalization constant ${\mathcal C}$ which reflects the choice of regulator in the field redefinition of the fermions.
To keep the time integrals finite, we have assumed that
$x_1$ is in the forward light cones of both $t_2$ and $t_3$. We have also dropped a trivial disconnected piece proportional to 
\be
\langle \phi(x_1)\rangle \langle \psi(t_2) \bar \psi(t_3) \rangle=
 {\mathcal C}\frac{ i\, h}{4 \pi |y_1| |t_2-t_3|^{\frac{g^2}{4 \pi^2}}} \Theta(t_2-t_3)\,.
\ee
Note that $\tanh^{-1} x = \tanh^{-1} x^{-1} \pm \frac{i \pi}{2}$.  Further $Q_0(x) = \tanh^{-1} x$ and $P_0(x) = 1$.  Thus we are
matching the form of the bulk-defect-defect three point function predicted earlier (\ref{eq:predictedthreepoint}) based solely on conformal invariance and the constraint from $\Box \phi = 0$. As $\ell_2 = 0 =\ell_3$, the sum collapses to 
terms with $\ell=0$ which may have $\Delta_0 = 0$ or 1.
In this case, the admixture of $P_0$ and $Q_0$ in (\ref{eq:3ptfinal}) corresponds to the $\Delta_0 = 1$ conformal block, coming from a defect operator expansion where the first term of $\phi(t,y)$ is the $\Delta_0 = 1$ operator $\phi(t, 0)$ (or possibly $\phi(t,0) \bar \psi(t) \psi(t)$).

\vskip 0.1in

We pause to consider the significance of this result.  The expression (\ref{eq:3ptfinal}) does not transform nicely under time reversal.  The hyperbolic tangent is odd under $\nu \to -\nu$, and the theta function $\Theta(t_2-t_3)$ is not even an eigenfunction under $t \to -t$.  This type of three point function is forbidden by the time reversal invariance assumption in \cite{Lauria:2020emq} that led to a restriction of the defect operator spectrum that in turn was crucial in proving triviality.
Another assumption in their proof, necessary for cluster decomposition, was that any operator with dimension zero must be the identity.  Here    again, this example evades the assumption as it possesses the dimension zero protected operator $\bar \psi \psi$.
We are thus in a situation where 
 the proof \cite{Lauria:2020emq} (that the operators in the DOE of $\phi$ are GFF's) is not applicable.
 Nevertheless, the theory is arguably still interacting as it has interesting three point functions with operators, $\psi$ and $\bar \psi$, that are not in the defect expansion of $\phi$ and which also have nontrivial anomalous dimension.

\vskip 0.1in

To gain further confidence in this result,
one can carry out a perturbative calculation.
The free
fermion two point function takes the form
\begin{equation}
\langle \psi(t) \bar \psi(0) \rangle = \Theta(t) \ .
\end{equation}
The free scalar field in 4d has the two point function
\begin{equation}
\langle \phi(x) \phi(0) \rangle = \frac{1}{4 \pi^2 x^2} \ .
\end{equation}
The three point function at leading order should follow from the integral, where as before we take $x_1$ to lie in the forward light cone of $t_2$ and $t_3$,
\begin{equation}
\langle \phi(t_1, y_1)  \psi(t_2) \bar \psi(t_3) \rangle = \frac{ig}{4\pi^2}  \int dt \, \frac{\Theta(t_2-t) \Theta(t-t_3) }{-(t_1-t)^2 + y_1^2} \ .
\end{equation}
The factor of $i$ comes from expanding out our Minkowski signature path integral to linear order in $g$.
The expression
 simplifies to
\begin{equation}
ig \Theta(t_2-t_3)
 \int_{t_3}^{t_2}
dt \frac{1}{-(t_1-t)^2 + y_1^2} = ig \Theta(t_2-t_3) \left. \frac{\tanh^{-1} \left( \frac{|y_1|}{t-t_1} \right)}{|y_1|} \right|_{t_3}^{t_2} \ ,
\end{equation} 
reproducing the previous expression but without the $|t_2-t_3|^{g^2/4 \pi^2}$ which evaluates to one at leading order in $g$.

\subsubsection*{Further three point functions}

We can also compute correlators for various classes of spinning defect operators, which must
also conform to the constraints imposed by conformal symmetry 
(\ref{eq:predictedthreepoint}).  We construct these spinning defect operators by dressing the fermions with factors $\phi$ and its derivatives on the defect.  Using the notation $(x_1 - t_2)^2 = -(t_1-t_2)^2 + y_1^2$,
for example we have
\begin{align}\label{eq:phis1s0}
\langle \phi(x_1) \der_i\phi\psi(t_2) \bar \psi(t_3) \rangle &= \frac{1}{2\pi^2} \frac{y_{1,i}}{(x_1-t_2)^4}
|t_2-t_3|^{-\frac{g^2}{4\pi^2}} \Theta(t_2-t_3)\ ,
\end{align}
setting the normalization constant ${\mathcal C}$ we had above to one.
This expression has $\ell_2 = 1$ and $\ell_3 = 0$ and reproduces (\ref{eq:predictedthreepoint}) with a single term in the sum, $\ell=1$.  
One can look at something more complicated:
\begin{align}\label{eq:phis1s1}
\langle \phi(x_1)\, \der_i\phi\psi(t_2) \,\der_j\phi\bar \psi(t_3) \rangle &= 
\frac{ig}{8\pi^4} \frac{\delta_{ij}}{|y_1||t_2-t_3|^{\frac{g^2}{4\pi^2}+4}}
\tanh^{-1} \left( \frac{1}{\nu} \right)
\Theta(t_2-t_3)
\end{align}
which could allow for up to three terms in the sum with  $\ell=2$, $1$ or $0$.  Only the $\ell=0$ term appears,
giving a result similar to the example without derivatives.
A result that leads to a simplification in the evaluation of both of these correlation functions is
\begin{align}
\left\langle \der_i \phi(t_2) \left( \int^{t_3} \phi(\tau) d\tau \right) \right\rangle &=0 \ .
\end{align}

As an extension of the correlator with two spin-one operators \eqref{eq:phis1s1}, the operators of higher transverse spins $(\ell_2,\ell_3\ge 1)$ can be evaluated in a similar manner, 
\begin{align}
\langle \phi(x_1)\,  (p_2 \cdot \der)^{\ell_2}\phi\psi(t_2) \, &(p_3\cdot\der)^{\ell_3}\phi\bar \psi(t_3) \rangle
=  \\
& \frac{ig}{16\pi^4} \frac{(-1)^{\ell_2+1}\delta_{\ell_2\,\ell_3}2^{\ell_2}\ell_2!\,(p_2\cdot p_3)^{\ell_2}}{|y_1||t_2-t_3|^{\frac{g^2}{4\pi^2}+2(\ell_2+1)}}
\tanh^{-1} \left( \frac{1}{\nu} \right)\Theta(t_2-t_3)\,, \nonumber
\end{align}
as the non-vanishing contribution  comes purely  from Wick contracting the two ``spinning" $\phi$'s. As with \eqref{eq:phis1s1}, only the $\ell = 0$ term appears. A similar extension of the correlator \eqref{eq:phis1s0} is
\be
\langle \phi(x_1) \,(p_2\cdot\der)^{\ell_2}\phi\psi(t_2)\, \bar \psi(t_3) \rangle = \frac{1}{4\pi^2} \frac{2^{\ell_2}\ell_2!\,(p_2\cdot y)^{\ell_2}}{(x_1-t_2)^{2(\ell_2+1)}}
|t_2-t_3|^{-\frac{g^2}{4\pi^2}} \Theta(t_2-t_3)\,,
\ee
and we see that only the $\ell=\ell_2$ term appears.

\vskip 0.1in

It is possible also to look at defect operators involving monomials of the form ${:}\phi^k \psi(t){:}$.
A complication here is that $\phi \psi = \partial_t \psi$ is a descendent operator.  More generally, a primary of this type will be a linear combination of monomials made from $\phi$, its time derivatives, and a single fermion.  For example, a scalar defect primary with dimension $\hat \Delta = 2 + \frac{g^2}{8 \pi^2}$ has the form 
\begin{equation}
\hat O = \left(\phi^2 + \frac{ig}{4\pi^2} \partial_t \phi \right) \psi \ .
\end{equation}
One finds
\begin{align}
\langle \phi(x_1) \hat O(t_2) \bar \psi(t_3) \rangle &= 
\frac{i g}{8 \pi^4} \frac{|y_1| \nu}{(x_1 - t_2)^4 |t_2-t_3|^{\frac{g^2}{4\pi^2}}} \Theta(t_2-t_3) \\
&= -\frac{ig}{16 \pi^4} 
\frac{\zeta^{1 + g^2/8\pi^2} Q_0^{2}(\nu)}{
(x_1-t_2)^{4 + g^2/4\pi^2} (x_1 - t_3)^{g^2/4\pi^2} |y_1|^{-1 - g^2/4 \pi^2}
}\Theta(t_2-t_3) 
\nonumber
\end{align}
which matches on the nose an $\ell=0$ contribution to the sum (\ref{eq:predictedthreepoint}) with $\Delta_0 = 1$.  We leave it as an exercise to the reader to verify that $\hat O$ has a vanishing two point function with
$\bar \psi$. 

\subsection{Stress Tensor, Displacement Operator, and $G$-function}

This section is not central to the main argument and can be skipped, 
but we would be remiss not to provide some of the central data of our defect CFT.
To wit, these are the one point function of the stress tensor, the two point function of the displacement operator, and the $G$-function.  We will see that  this data is not sensitive to the marginal coupling $g$.

\vskip 0.1in

Consider the improved stress tensor for the bulk scalar:
\begin{equation}
T^{\mu\nu} = (\partial_\mu \phi) (\partial_\nu \phi) - \frac{\eta_{\mu\nu}}{2} (\partial \phi)^2 - \xi (\partial_\mu \partial_\nu - \eta_{\mu\nu} \partial^2 ) \phi^2 \ .
\end{equation}
In the conformal case $\xi = \frac{d-2}{4 (d-1)}$. 
The first object of interest, the expectation value of $\langle T^{\mu\nu} \rangle=0$ vanishes in the background (\ref{eq:phibg}).

\vskip 0.1in

The divergence of the proposed stress tensor is
\begin{equation}
\partial_\mu T^{\mu\lambda}(x)  = (\partial^\lambda \phi) \Box \phi(x) = - (\partial^\lambda \phi) (g \bar \psi \psi + h) \delta^{(d-1)}(x^i) \ .
\end{equation}
We don't expect conservation for spatial indices $\lambda=i$ because the presence of the defect breaks translation symmetry in these directions.  However, we do expect it for $\lambda = 0$.  Thus we require an improvement term for the stress tensor, coming from the defect fermions.  The obvious candidate is
\begin{equation}
T^{\mu\nu} \to T^{\mu\nu}   +  \phi(g  \bar \psi \psi +h)\delta^{(d-1)}(x^i) \eta^{0 \mu} \eta^{0 \nu} \ .
\end{equation}
This choice requires the separate conservation of the boundary current, $\partial_0 (\bar \psi \psi) = 0$, which follows from the fermion equations of motion.

\vskip 0.1in

The trace of the stress tensor on the other hand is now
\[
T^\mu_\mu(x) 
= \phi \left(\frac{d-2}{2} \Box \phi + (g  \bar \psi \psi +h)\delta^{(d-1)}(x^i) \right) \ .
\]
In precisely four dimensions, this trace vanishes by the equation of motion for $\phi$.
The analysis here is purely classical. From a quantum perspective, this trace should be corrected by the beta functions for $g$ and $h$.  However, as we saw previously, these beta functions vanish in $d=4$, and the naive classical analysis remains valid.
(The more perhaps familiar situation is to find a vanishing trace for a choice of dimension $d=4-\epsilon < 4$ where the beta function vanishes, as happens for example with the Wilson-Fisher fixed point \cite{Brown:1979pq,Paulos:2015jfa,Ge:2025fsm}.)

\vskip 0.1in

From the divergence of the stress tensor, we identify the displacement operator:
\begin{equation}
D^i = -  (\partial^i \phi) (g \bar \psi \psi +h) \ .
\end{equation}
The (connected portion of the) displacement operator two point function is (in $d=4$)
\begin{equation}
\langle D^i(t) D^j(t') \rangle = h^2 \lim_{x,y \to 0} \partial^i_x \partial^j_y \langle \phi(x,t) \phi(y,t') \rangle =  \frac{h^2}{2 \pi^2} \frac{ \delta^{ij}}{(t-t')^4} \ .
\end{equation}
We can compute this operator exactly, using the rescaling trick $\psi = e^{i g \int \phi dt} \psi'$.  The fermionic correlation functions vanish, proportional to $\Theta(t) \Theta(-t)$.   The conclusion is that
the displacement operator has nonzero norm only if $h \neq 0$.  
(As the displacement operator two point function is unaffected by the $g \phi \bar \psi \psi$ coupling, the results here duplicate the discussion in section 5.4 of \cite{Billo:2016cpy}.)

\vskip 0.1in

The three point function of $D^i$ clearly vanishes because $\langle \phi(x) \phi(y) \phi(z) \rangle$ vanishes.  
There should be no tilt operator associated with the $\psi \to e^{i g \alpha_0} \psi$ symmetry as this symmetry is not broken and is anyway a purely boundary symmetry.  
On the other hand, there is a tilt operator 
$T = g \bar \psi \psi + h$ associated with the bulk shift symmetry, whose two point function is the number $h^2$.  

\vskip 0.1in

We now calculate the $G$-function  for our model.
Using Euclidean signature here, we follow an established perturbative strategy
\cite{Klebanov:2011gs,Cuomo:2021kfm,Ge:2024hei}. The $G$-function is defined as the logarithm of the ratio between the defect partition function $\mZ^{(g,h)}$ on the sphere $S^d$ of radius $R$ and the free scalar partition function $\mZ^{\phi}$ on the sphere $S^d$, as well as the free fermion partition function $\mZ^{\psi}$ on a great circle $S^1$ of the sphere $S^d$
\be
\ln G \equiv \ln \left( \frac{\mZ^{(g,h)}_{S^d}}{ \mZ^{\phi}_{S^{d}} \mZ^{\psi}_{S^1}} \right)\,.
\ee
The field redefinition \eqref{eq:trivialization} (after an adaptation to the Euclidean case) leaves the fermionic functional measure invariant. This factors out the ``free'' fermion partition function from the full defect one $\mZ^{(g,h)}_{S^d} = \mZ^{(0,h)}_{S^d} \mZ^{\psi}_{S^1}$, making the $G$-function independent of  the Yukawa coupling $g$. 
After canceling the free fermion part, the evaluation of the $G$-function simplifies,
\be
\ln G = \ln \left( \sum_{l} \frac{(-h)^{2l}}{2^l l!}\left(\int \sqrt{\gamma_{1}}\sqrt{\gamma_{2}} \, d\tau_1 d\tau_2 \,
\langle \phi(\tau_1) \phi(\tau_2) \rangle
\right)^l
\right) \stackrel{d=4}{=} \ln 1 =  0\,.
\ee
The result follows from the
 integrated two-point function on the great circle (of radius $R$), using the stereographic coordinates $ds^2_{S^1} = \frac{4R^2 d\tau^2}{(1+\tau^2)^2}$, where the range of $\tau$ is the whole real line,
\begin{align}
&~~~~\int \sqrt{\gamma_{1}}\sqrt{\gamma_{2}} \, d\tau_1 d\tau_2 \,
\langle \phi(\tau_1) \phi(\tau_2) \rangle 
= \Cpp\int d\tau_1 d\tau_2 \frac{\left( \frac{2R}{1+\tau_1^2} \right)^{1-\Delta} \left( \frac{2R}{1+\tau_2^2} \right)^{1-\Delta}}{(\tau_1 - \tau_2)^{2\Delta}}\no\\
&= \Cpp\frac{(2R)^{2-2\Delta}\pi ^{3/2} \Gamma \left(\frac{1}{2}-\Delta
   \right)}{\Gamma (1-\Delta )}
= \frac{(2R)^{4-d}\pi ^{\frac{3}{2}-\frac{d}{2}} \Gamma
   \left(\frac{3}{2}-\frac{d}{2}\right) \Gamma
   \left(\frac{d}{2}-1\right)}{4 \Gamma
   \left(2-\frac{d}{2}\right)}\,.
\end{align}
Using dimensional regularization,
the result vanishes for all even $d \geq 4$ and diverges for all odd $d \geq 3$.
While the IR behavior of the integral is finite due to the mass of the scalar on the sphere, there is a UV divergence from the coincident limit $\tau_2 \to \tau_3$ in all $d \geq 3$.  Dimensional  regularization, however, is only sensitive to log divergences, which appear only in odd dimensions $d \geq 3$.
Regardless of the divergence structure, the $g$-function
should come from the finite term, which is absent in $d=4$, indicating $\ln G=0$.

\section{Discussion}\label{sec:discuss}

In introducing our fermionic toy model, we mentioned in passing the related Bose-Kondo type models of refs.\ \cite{Cuomo:2022xgw,Komargodski:2025jbu}.
We should make some remarks about their status as additional ``non-trivial'' ({\it \`a la} ref.\ \cite{Lauria:2020emq}) examples of line defects.  
In this Bose-Kondo set-up, our fermions $\psi$ and $\bar \psi$ are traded for a pair of bosonic raising and lowering oscillators $z_1$, $z_2$, $z_1^\dagger$ and $z_2^\dagger$ which additionally satisfy a constraint $ z^\dagger z = s$, that freezes the total excitation level of the defect.  Further our single scalar field is traded for a free O(3) model along with defect interactions controlled by the Pauli matrices $z^\dagger \sigma^a z \phi_a$.  With these more complicated interactions, the theory can no longer be solved by a simple field redefinition.  Nevertheless, fixed points can be found in a $4-\epsilon$ expansion (although the fixed point may not persist all the way down to $\epsilon=1$) \cite{Cuomo:2022xgw}.   That fixed points exist is consistent with a lack of time reversal invariance of the correlation functions.  The natural action of time reversal in this theory sends $z_1 \to z_2^\dagger$ and $z_2 \to - \bar z_1^\dagger$ \cite{Komargodski:2025jbu}.  Thus defect-defect-bulk three point functions involving the $z$ would naturally use the cross ratio $\nu$ instead of $\xi$.  

\vskip 0.1in

While interesting and arguably more closely connected to the real world 
\cite{Vojta1999} 
than our toy model, unfortunately these Bose-Kondo models
do not provide as clean a counter-example as one would like.  The proof of triviality \cite{Lauria:2020emq} relies heavily on unitarity -- in particular the insistence that the operator scaling dimensions are bounded below -- which is lost for non-integer $\epsilon$ \cite{Hogervorst:2015akt}.  In constrast, our toy model is conformal exactly in $d=4$ dimensions.  

\vskip 0.1in

There is an obvious extension of our work to line defects in free Maxwell theory.
Indeed, 
we began working on this project with the mistaken impression that the extension
of the theorem 
of ref.\ \cite{Lauria:2020emq} to parity breaking line defects with bulk scalars would be straightforward, that the next logical step was to prove line defects in Maxwell theory have a decoupled GFF sector.
Previously,  one of us had established the existence of such a sector for $d=1+1$ dimensional surface defects in free Maxwell \cite{Herzog:2022jqv}.  Moreover, just as free scalars with a boundary admit a rich class of conformal theories \cite{Behan:2020nsf,Behan:2021tcn}, free Maxwell in $d=3+1$ dimensions is known to admit a rich class of conformal theories (see for example \cite{DiPietro:2019hqe}).
Thus it seemed likely that what was true for the massless free scalar would be true for free Maxwell.

\vskip 0.1in

Extending section \ref{sec:correlations} to write down the allowed form of
$\langle F_{\mu\nu}(x) \hat O_{\ell}(t) \hat O_{\ell'}(t') \rangle$ is straightforward.  Building on established formalism  \cite{Herzog:2020bqw,Herzog:2022jqv} and the polarization structures discussed in appendix \ref{app:polarization}, one can find
tensor structures with the proper transformation properties.  Indeed,
starting with the cross ratio $\nu$ (\ref{eq:nucrossratio}), 
the non-physical singularities that were key in the proof of
\cite{Lauria:2020emq} appear to be absent.

\vskip 0.1in

One would like then to find a simple model of a nontrivial line defect in free Maxwell.  
What goes wrong with trading the free massless scalar field $\phi$
used in the toy model in section \ref{sec:toymodel} for a free Maxwell field?
The problem is that there is now an honest gauge symmetry in the model.  
It is only the gauge invariant correlation functions that must respect the 
conformal symmetry.  Gauge non-invariant correlators respect conformal symmetry only up to gauge transformation.
The fermionic field $\psi$ is not gauge invariant on its own. 
The nonlinear transformation that we used to trivialize the scalar theory now
has the interpretation of a fermionic field dressed by a Wilson line
\[
\exp \left(- i g \int^t A_\tau d \tau \right) \psi(t)  \ .
\] 
It is this dressed field which is gauge invariant.  However, as we saw in the case
of the scalar field, this dressing has the effect of producing a free fermion.  
The gauge invariant correlation functions in this defect Maxwell field theory then  decouple.  One is left with the direct product of a free Maxwell theory in $d=3+1$ and a free fermion in $d=1$.  

\vskip 0.1in

While our scalar model does not generalize in a particularly interesting way to the Maxwell case,
the bulk-defect-defect three point functions leave room for a more interesting interacting line defect.  
It will be interesting to see if an
interacting line defect in $d=3+1$ Maxwell can be developed.  
It would also be worthwhile to search for more examples of interacting line defects with free bulk scalars or free bulk fermions, perhaps in dimensions $d \neq 4$.  
One particularly simple generalization is to introduce additional fermionic flavor degrees of freedom on the line defect, possibly also with quartic interactions.
An extension of our studies on the interacting line defects to the finite temperature case looks interesting and promising as well \cite{Marchetto:2023fcw,Barrat:2024aoa}.
We leave these tasks for future work.

\vskip 0.1in

{\bf Note Added:}  We should mention that the
non-physical singularity in the bulk-defect-defect three point function has been exploited in 
higher dimensions $p>1$ to put ``locality constraints'' on  quantum field theories in AdS$_{p+1}$ \cite{Levine:2023ywq,Levine:2024wqn}.  From this perspective, the case analyzed here, through Weyl transformation, is that of a conformally coupled scalar field in AdS$_2$. About a month after the initial version of this manuscript appeared on the arXiv, 
ref.\ \cite{Loparco:2025aag} explored how the same lack
of parity symmetry discussed here can relax this ``locality constraint'' for more general QFTs in AdS$_2$.


\section*{Acknowledgment}
We would like to thank Vladimir Schaub for collaborations at the initial stages of the work. We also thank Dio Anninos, Christian Copetti, Masazumi Honda, Anatoly Konechny, Petr Kravchuk, Neil Lambert, Mark Mezei, and Miguel Paulos for useful discussions. 
We would also like to thank Edo Lauria, Jacopo Sisti, and Balt van Rees for comments on the manuscript.
SB and CH thank the Isaac Newton Institute for Mathematical Sciences for support and hospitality during the program ``QFT with BIDS" when part of this work was undertaken.
DG thanks the Theoretical Physics Group of King's College London for hospitality where this work was started. 
DG is supported in part by the JSPS Grant-in-Aid for Transformative Research Areas (A) “Extreme Universe” No.\ 21H05182 and No.\ 21H05190. 
SB and CH are supported in part by the STFC under grant ST/X000753/1 and EPSRC under grant EP/Z000580/1.


\appendix

\section{Cross Ratios}
\label{app:nu}

\begin{figure}
\begin{center}
a) \includegraphics[width=2.5in]{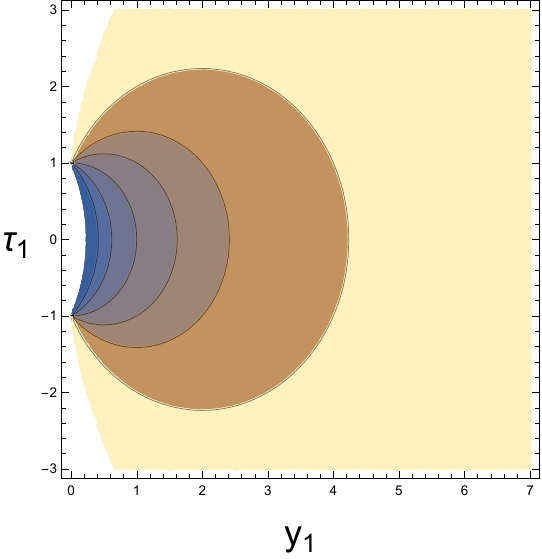}
b) \includegraphics[width=2.5in]{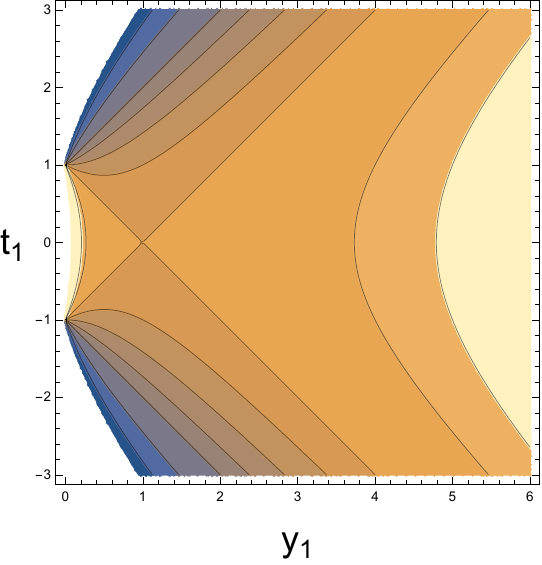}
\end{center}
\caption{
Contour plots for a) the Euclidean cross ratio $\nu_E$ when $(\tau_2, \tau_3) = (1,-1)$ and b) the Minkowski cross ratio $\nu$ when $(t_2, t_3) = (1,-1)$.
Darker colors are more negative.  Lighter colors are more positive.
\label{fig:nuplots}
}
\end{figure}

Here, we include a more detailed discussion about the cross ratio $\nu$. 
The Euclidean version of the cross ratio is simpler to understand.  Freezing $\tau_2$ and $\tau_3$, we can look at constant contours in the $(\tau_1, y_1)$ plane
(see figure \ref{fig:nuplots}a):
\begin{equation}
\left( y_1 - \nu_E \frac{\tau_2-\tau_3}{2} \right)^2 + \left( \tau_1 - \frac{\tau_2 + 
\tau_3}{2} \right)^2 =\frac{1}{4} (\tau_2-\tau_3)^2(\nu_E^2+1)  \ .
\end{equation}
These are circles of radius $\frac{|\tau_2 - \tau_3|\sqrt{\nu_E^2+1}}{2}$ that pass through the points $(\tau_2, 0)$ and $(\tau_3, 0)$.  One subtlety is that $y_1$, interpreted as a distance from the defect, should be positive.  When $\nu_E = 0$, the allowed set of $(\tau_1, y_1)$ with $y_1 > 0$ is a semicircle.   For $\nu_E > 0$
and $\tau_2 > \tau_3$, the contour lies outside this semicircle.  For $\nu_E < 0$ and $\tau_2 > \tau_3$ in contrast, the contour lies inside this semicircle.

\vskip 0.1in

Analytically continuing to Lorentzian signature, the constant $\nu$ contours are hyperbolae
(see figure \ref{fig:nuplots}b),
\begin{equation}
\left( y_1 - \nu \frac{t_2-t_3}{2} \right)^2 - \left( t_1 - \frac{t_2 + 
t_3}{2} \right)^2 =\frac{1}{4} (t_2-t_3)^2(\nu^2-1) \ ,
\end{equation}
whose orientation and placement depend on where $\nu$ sits on the numberline compared with $0$, $1$, and $-1$.  
The values $\nu = \pm 1$ are special because here $(y_1, t_1)$ lie on the forward or backward lightcones of
$(0, t_2)$ and $(0, t_3)$.  Assuming $t_2>t_3$, $\nu$ gets very large and positive at the origin and as $\lim_{y_1 \to \infty} (y_1,0)$. 
Along the $t_1$ axis, in the limits $t_1 \to \pm \infty$, $\nu$ gets very large and negative.

\vskip 0.1in

\begin{figure}
    \centering
    \includegraphics[width=0.5\linewidth]{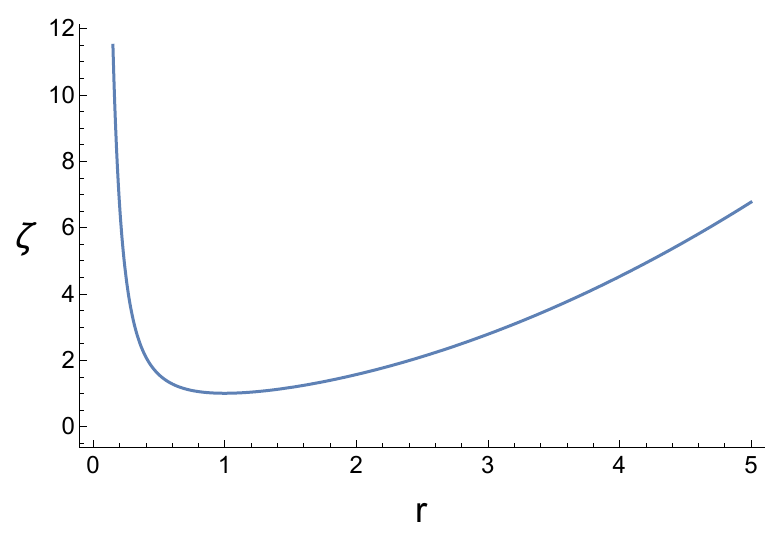}
    \caption{Cross ratio $\zeta$ in the conformal frame as a function of transverse distance $r$.}
    \label{fig: zeta vs r in the conformal frame}
\end{figure}

Given the two-to-one mapping $\zeta = \nu^2 - 1$
(or equivalently in Euclidean signature $\zeta_E = 1+ \nu_E^2$, with the identification $\zeta_E = - \zeta$),
the standard $\zeta$ cross ratio (\ref{eq: definition of zeta cross ratio}) provides a coarser parametrization of space-time.
Let us consider the behavior of $\zeta_E$
in a specific conformal frame. 
We set $x_2 = (1, 0, \vec{0})$, $x_3 = (-1, 0, \vec{0})$, and
$x_1 = (0, r, \vec{0})$.  Such a coordinate choice can always be achieved by first using a special conformal transformation to put $t_1$ midway between $t_2$ and $t_3$,  a translation to set $t_1 = 0$, a dilatation to make $|t_2| = |t_3| = 1$, and then a rotation to align $x_1$ with a coordinate axis, 
leaving the single parameter $r = |y_1|$.
The cross ratio in this frame is
\begin{equation}
    \label{eq: zeta in the conformal frame}
    \zeta_E = \frac{(1+r^2)^2}{4r^2},
\end{equation}
fully determined by the location of $x_1$, plotted in figure \ref{fig: zeta vs r in the conformal frame}.
Clearly, $\zeta_E$ is not an injective function of the transverse distance $r$, as it is symmetric under inversion $r \to \frac{1}{r}$.  It is minimized at $r=1$, a fixed point under inversion, corresponding to putting $x_1$ on the unit sphere. While at other locations, inversion symmetry makes $\zeta_E$ as a cross ratio  unable to distinguish whether a point falls within or outside of the unit sphere.

\vskip 0.1in

It is simple to manipulate $\zeta_E$ in a piece-wise manner to construct a new cross ratio that is invertible over its entire domain. In general, such a cross ratio will not be smooth. It is a non-trivial result that the cross ratio $\nu_E$ is smooth, special to $p=1$. For more general $p$, the construction will fail.

\vskip 0.1in

We examine the group transformation properties of $\nu_E$ in more detail.  The cross ratio is invariant under
\begin{equation}
\tau_1 \to \frac{-\tau_1}{\tau_1^2 + y_1^2} \ , \; \; \;
y_1 \to \frac{y_1}{\tau_1^2 + y_1^2} \ , \; \; \;
\tau_2 \to -\frac{1}{\tau_2} \ , \; \; \;
\tau_3 \to -\frac{1}{\tau_3} \ ,
\label{eq:signinversion}
\end{equation}
which is a member of the $SO^+(1,2) = PSL(2, {\mathbb R})$ group of transformations.  
Indeed, in the limit $y_1 \to 0$, we see that this transformation
reduces to a particular M\"obius transformation, $\tau_i \to -1 /\tau_i$ where the general set of transformations is 
$\tau \to \frac{a \tau + b}{c\tau + d}$ with $ad - bc = 1$.
However, the ratio $\nu_E$ 
picks up a sign under time reversal $\tau_i \to - \tau_i$, which is not a M\"obius transformation with determinant one.
Give this sign reversal, there must also be a sign reversal under the transformation usually called inversion
\begin{equation}
\label{eq:inversion}
    \tau_1 \to \frac{\tau_1}{\tau_1^2 + y_1^2} \ , \; \; \;
y_1 \to \frac{y_1}{\tau_1^2 + y_1^2} \ , \; \; \;
\tau_2 \to \frac{1}{\tau_2} \ , \; \; \;
\tau_3 \to \frac{1}{\tau_3} \ .
\end{equation}
Thus we see in a completely explicit fashion that $\nu_E$ picks up a sign under inversion.

\vskip 0.1in

Exploring these group transformations in Lorentzian signature is more subtle.  The inversion transformation (\ref{eq:inversion}) continues to
\begin{equation}
t \to \frac{t}{-t^2 + y^2} \ , \; \; \;
y \to \frac{y}{-t^2 + y^2} \ .
\end{equation}
Being careful to take the appropriate branch cut of
$\sqrt{-t_1^2 + y_1^2}$, $\nu$ picks up a minus sign.

\vskip 0.1in

A natural question to ask is whether this cross ratio $\nu$ could be adjusted to be invariant under other choices of discrete group action.  For example, one might consider taking the absolute value of $|t_2-t_3|$ in the denominator.  Such an absolute value, however, has the unfortunate consequence of spoiling invariance under $PSL(2, {\mathbb R})$ because of the extra sign now picked up under the transformation (\ref{eq:signinversion}).

\section{Polarization Tensors}
\label{app:polarization}

We must be able to decompose the bulk-defect-defect three point functions into representations of the transverse rotation group $SO(d-1)$.  If the two defect operators transform in traceless symmetric representations $\ell_1$ and $\ell_2$, then the three-point function should decompose into traceless symmetric representations with a number of indices that ranges from $|\ell_1 - \ell_2|$ to $\ell_1 + \ell_2$.

\vskip 0.1in

To encode these representations, we introduce two light-like vectors $p_1$ and $p_2$ along with the unit normal vector $n = y_3/|y_3|$. 
 The polarization tensors in the correlation function must be polynomials in
 the words
\begin{equation}
w_{abc} = (n \cdot p_1)^a (n \cdot p_2)^b (p_1 \cdot p_2)^c
\end{equation}
for $a$, $b$, and $c$ non-negative 
integers.
Moreover, we demand these polynomials be eigenfunctions of the total angular momentum $L^2 = (L_1 + L_2)^2$ acting on $p_1$ and $p_2$.

\vskip 0.1in

We find that
\begin{eqnarray}
\label{generalbasicresult}
L^2 w_{abc} =  (a+b)(a+b+d-3)  w_{abc} - 2a b  \, w_{a-1, b-1, c+1} \ , 
\end{eqnarray}
which points the way toward a recursion relation for these polynomials. 

\vskip 0.1in

Indeed, consider a polynomial of the form
\[
P_{abc} \equiv c_b w_{abc} + c_{b-1} w_{a-1, b-1, c+1} + \ldots + c_{0} w_{a-b,0,c+b} \ ,
\]
assuming without loss of generality that $a>b$.
Introducing 
\[
\chi \equiv \frac{(n \cdot p_1)(n \cdot p_2)}{p_1 \cdot p_2} \ ,
\]
we can rewrite the polynomial in the form
\[
P_{abc} \equiv (n \cdot p_1)^{a-b} (p_1 \cdot p_2)^{c+b} p_{a,b} (\chi) \ ,
\]
and where
\[
p_{a,b}(\chi) = c_0 + c_1 \chi + c_2 \chi^2 + \ldots + c_b \chi^{b} \ .
\]

\vskip 0.1in

In this form, the condition that $P_{abc}$ is an eigenfunction of $L^2$  implies the following recursion
relation on the $c_n$:
\[
c_{n+1} = - \frac{(b-n)(d-3+2a+2n)}{(n+1)(n+1+a-b)} c_n  \ .
\]
The eigenvalue $(a+b)(a+b+d-3)$ is set by the action of $L^2$ on the $c_b$ term in the polynomial, which has no $c_{b+1}$ term to counterbalance it.
This recursion relation is solved by
\begin{equation}
c_n = \frac{2^n}{n!}
\frac{(-b)_n \left( a + \frac{d-3}{2} \right)_n}
{(a-b+1)_n} c_0 \ ,
\end{equation}
which sums to a hypergeometric polynomial:
\begin{equation} \label{eq:PolyOri}
P_{abc} = c_0\, w_{a-b,0,c+b}\, {}_2 F_1
\left( a + \frac{d-3}{2}, -b; a-b+1; 2 \chi \right) \ .
\end{equation}
These hypergeometric functions reproduce what we found for the functions $h_\ell(\chi)$ (\ref{eq:hchi})
from enforcing that the bulk-defect-defect three point function satisfied the free equation of motion for the scalar field.  We just need to make the replacements
\begin{equation}
a = \frac{\ell + \ell_2 - \ell_3}{2} \ , \; \; \;
b = \frac{\ell - \ell_2 + \ell_3}{2} \ .
\end{equation}

\paragraph{An alternative structure:}
In fact, there is another structure for the polynomials in $d=4$, constructed using the three-dimensional Levi-Civita symbol, with words given as
\be
\bar w_{abc} = (n \cdot p_1)^a (n \cdot p_2)^b (p_1 \cdot p_2)^{c-1} (\vareps_{ijk}\, p_1^i\, p_2^j\, n^k) \,.
\ee
Under the action of the total angular momentum $L^2$,
\be
L^2 \bar w_{abc} = (a+b+1)(a+b+2) \bar w_{abc} - 2 a b \bar w_{a-1,b-1,c+1}\,.
\ee
Similarly, a polynomial can be constructed using these words
\be
\bar P_{abc} \equiv \bar c_b \bar w_{abc} + \bar c_{b-1} \bar w_{a-1, b-1, c+1} + \ldots + \bar c_{0} \bar w_{a-b,0,c+b}\,.
\ee
Requiring the polynomial being an eigenfunction of $L^2$ gives a recursion relation on its coefficients
\be
\bar c_{n+1} = -\frac{(b-n )(2 a+2 n+3) }{(n+1) (a-b+n+1)} \bar c_n\,.
\ee
 Solving the recursion relation gives
\be
\bar c_n = \frac{2^n}{n!} \frac{\left(a+\frac{3}{2}\right)_n \left(-b\right)_n }{\left(1+ a-b\right)_n } \bar c_0\,,
\ee
Then 
\be\label{eq:PolyExtra}
\bar P_{abc} = \bar c_0\, \bar w_{a-b,0,c+b}\, _2F_1\left(a+\frac{3}{2},-b;a-b+1;2 \chi\right)\,,
\ee
where the above hypergeometric function is related to the Jacobi polynomial
\be
_2F_1\left(a+\frac{3}{2},-b;a-b+1;2 \chi\right)
=\frac{b!}{(a-b+1)_b} P_b^{\left(a-b,\frac{1}{2}\right)}(1-4 \chi)\,.
\ee

\section{Bulk-Defect-Defect Correlations from the DOE}
\label{app:bulkdefectdefect}
Using the DOE \eqref{eq:phidOPE}, we provide an alternative approach of calculating the bulk-defect-defect three point function. Expanding the bulk operator gives
\begin{equation}
\langle \phi(x_1) \hat O_{\ell_2} (t_2) \hat O_{\ell_3}(t_3) \rangle = 
\sum_{\hat O_{\ell}} \frac{\mB_{\phi \hat O{_\ell}}}{\mN_{\hat O_\ell \hat O_\ell}} 
\frac{1}{|y_1|^{\Delta - \Delta_{\ell}}}\mD^{\Delta_{\ell}} (|y_1|^2 \der_{t_1}^2) \left\langle
\hat O^\ell(t_1) \hat O_{\ell_2} (t_2) \hat O_{\ell_3}(t_3)
\right\rangle
\,,
\end{equation}
noting $\hat O^\ell$ is contracted with $n_1$ while $\hat O_{\ell_{2,3}}$ are contracted with the lightlike vectors $p_{2,3}$. The non-vanishing channels are labeled by $\ell \in \{|\ell_2-\ell_3|,|\ell_2-\ell_3|+1,\dots,\ell_2 +\ell_3\}$. Without loss of generality, we assume $\ell_2>\ell_3$ for the following. This three-point function on the line has the following structure
\be\label{eq:defect3pt}
\left\langle
\hat O^\ell(t_1) \hat O_{\ell_2} (t_2) \hat O_{\ell_3}(t_3)
\right\rangle
=\frac{\mC_{ \hat O_{\ell_2} \hat O_{\ell_3} \hat O_\ell} \mP_{\ell_2, \ell_3}^\ell(n_1,p_2,p_3)}{|t_1-t_2|^{\Delta_3} |t_1 - t_3|^{\Delta_2} |t_2 -t_3|^{\Delta_1}}\,, 
\ee
with $\Delta_1 = \Delta_{\ell_2} + \Delta_{\ell_3} - \Delta_{\ell}$, $\Delta_2 = \Delta_\ell + \Delta_{\ell_3} - \Delta_{\ell_2} $ and $\Delta_3 = \Delta_\ell + \Delta_{\ell_2} - \Delta_{\ell_3} $; $\mC_{ \hat O_{\ell_2} \hat O_{\ell_3} \hat O_\ell}$ is the defect OPE coefficient,  $\mP_{\ell_2, \ell_3}^\ell(n_1,p_2,p_3)$ is an $SO(d-1)$ invariant polynomial constructed in terms of $n_1$ and $p_{2,3}$. They are identified up to a rescaling of the polynomials constructed in \eqref{eq:PolyOri} and \eqref{eq:PolyExtra}. More explicitly, in $d=4$ spacetime dimensions, for $\ell = \ell_2+\ell_3 -2m\,(m=0,\cdots,\ell_3)$
\be
\mP_{\ell_2, \ell_3}^\ell = \frac{
  (\ell_2-\ell_3+1)_{\ell_3-m} \,
   w_{\ell_2-\ell_3,0,\ell_3}}{(-2)^{\ell_3-m}\left(\ell_2-m+\frac{
  1}{2}\right)_{\ell_3-m}} \,
  _2F_1\left(\ell_2-m+\frac{1}{2},-\ell_3+m;\ell_2 -\ell_3+1;2 \chi\right)\,,
\ee
for $\ell = \ell_2+\ell_3 -2m+1\,(m=1,\cdots,\ell_3)$ in $d=4$ spacetime dimensions,
\be
\mP_{\ell_2, \ell_3}^\ell = \frac{
  (\ell_2-\ell_3+1)_{\ell_3-m} \,
  \bar w_{\ell_2-\ell_3,0,\ell_3}}{(-2)^{\ell_3-m}\left(\ell_2-m+\frac{
  3}{2}\right)_{\ell_3-m}} \,
  _2F_1\left(\ell_2-m+\frac{3}{2},-\ell_3+m;\ell_2 -\ell_3+1;2 \chi\right)\,.
\ee
The differential operator $\mD^{\Delta_\ell} (y^2_1 \der_{t_1}^2)$ acts only on the denominator of the defect three-point function \eqref{eq:defect3pt}, and has a compact form after the infinite sums,
\begin{align}
&~~~\mD^{\Delta_\ell} (y_1^2 \der_{t_1}^2) \frac{1}{|t_1- t_2|^{\Delta_3}|t_1-t_3|^{\Delta_2}}
= | x_{12} |^{-\Delta_3}| x_{13} |^{-\Delta_2} 
   \, _2F_1 \left( \frac{\Delta_3}{2}, \frac{\Delta_2}{2}, \Delta_\ell+\frac{1}{2};-\frac{1}{\zeta} \right)\,.
\end{align}
 A derivation of this compact form  can be found in Appendix C.2 of \cite{Okuyama:2023fge}, where the analysis was done in the momentum space of the bulk operator and Fourier transformed back to the position space in the end.  Combining all these pieces together, the bulk-defect-defect three-point function 
is
\be
\langle \phi(x_1) \hat O_{\ell_2} (t_2) \hat O_{\ell_3}(t_3) \rangle = 
\sum_{\hat O_{\ell}} \frac{\mB_{\phi \hat O{_\ell}} \mC_{ \hat O_{\ell_2} \hat O_{\ell_3} \hat O_\ell}  }{\mN_{\hat O_\ell \hat O_\ell}} 
\frac{\mP_{\ell_2, \ell_3}^\ell(n_1,p_2,p_3)\, \, _2F_1 \left( \frac{\Delta_3}{2}, \frac{\Delta_2}{2}, \Delta_\ell+\frac{1}{2};-\frac{1}{\zeta} \right)}{|y_1|^{\Delta - \Delta_{\ell}}| x_{12} |^{\Delta_3}| x_{13} |^{\Delta_2} |t_{23}|^{\Delta_1}}\,.
\ee

\section{Beta Function}
\label{app:beta}

Our Feynman rules are as follows.
The propagator for the fermion is
\begin{equation}
\frac{i}{\omega + i \varepsilon} \ .
\end{equation}
The propagator for the scalar $\phi$ is
\begin{equation}
\frac{-i}{k^2 +m^2 - i \varepsilon} \ .
\end{equation}
The vertex is $i g$.  Assembling these factors for a given diagram will then give the amplitude with an extra factor of the square root of minus one, $i T$.  
We include a wave function renormalization $Z_\psi \psi^\dagger (i \partial_t) \psi$ and a vertex renormalization $g Z_g \phi \bar \psi \psi$ in the Lagrangian.

\subsection{One Loop Corrections}

\begin{figure}
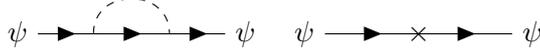

\begin{center}
\proponeloop
\proponeloopCT
\end{center}
\caption{One loop self energy and counterterm.}
\label{fig:proponeloop}
\end{figure}

The one loop correction to the fermion propagator (see figure \ref{fig:proponeloop}) is
\begin{equation}
i \Pi_1(\mu)  = (i g)^2  \int \frac{d \omega \, d^{d-1} p_\perp}{(2\pi)^d} \frac{-i}{-\omega^2 + p_\perp^2 - i \varepsilon} \frac{i}{\omega + \mu+ i \varepsilon}
\end{equation}
which evaluates to 
\begin{eqnarray}
i \Pi_1(\mu) &=&\frac{ g^2 \Vol(S^{d-2})}{(2\pi)^d} \frac{\pi^2}{(-i \mu)^3}((-i \mu)^d + (i \mu)^d) \csc(\pi d)^2 \sin\left(\frac{\pi d}{2} \right) \nonumber \\
&=&i  \mu g^2 \left( \frac{1}{4 \pi^2 \epsilon} - \frac{1}{8 \pi^2} \left( \log \left(\frac{\mu^2}{4 \pi} \right)  - \psi\left(\frac{3}{2}\right) \right) + O(\epsilon) \right) \ ,
\end{eqnarray}
where in the last line we expanded in the limit $\epsilon \ll 1$, with $d=4-\epsilon$.
Renormalization demands  $i \Pi_1(\mu) + i (Z_\psi - 1) \mu$ to be finite, which implies 

\begin{equation}
Z_\psi =  1 - \frac{g^2}{4\pi^2 \epsilon}  \ .
\end{equation}

\begin{figure}
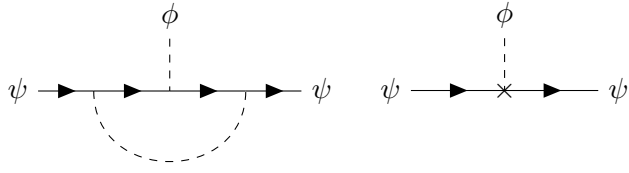

\begin{center}
\vertexoneloop
\raisebox{2.5em}{
\vertexoneloopCT
}
\end{center}
\caption{One-loop vertex and counterterm.}
\label{fig:vertexoneloop}
\end{figure}

The one loop vertex correction (see figure \ref{fig:vertexoneloop}) is
\begin{equation}
iV(\mu_1, \mu_2) = (i g)^3 \int \frac{d \omega \, d^{d-1} p_{\perp}}{(2\pi)^d} \frac{-i}{-\omega^2 + p_\perp^2} \frac{i}{\omega + \mu_1} 
\frac{i}{\omega+\mu_2}
\end{equation}
which evaluates to
\begin{equation}
i V(\mu_1, \mu_2) = i g^3 \left( \frac{1}{4 \pi^2 \epsilon}  + \frac{1}{8 \pi^ 2} \left( \log 4 \pi + \psi\left(\frac{3}{2}\right) + \frac{ \mu_2 \log \mu_2^2 - \mu_1 \log \mu_1^2 }{\mu_1 - \mu_2} \right) + O(1) \right)  \ .
\end{equation}
At one loop, we need
\[
i g Z_g  + \frac{ig^3}{4 \pi^2 \epsilon}
\]
to be finite, which implies 
\begin{equation}
Z_g = 1 - \frac{g^2}{4 \pi^2 \epsilon} \ .
\end{equation}
As demanded by the Ward identity, there is an equality $Z_\psi=Z_g$.\footnote{%
 Ref.\
 \cite{Bashmakov:2024suh} considered many defect models in free theories, of which this example was one.
 In this case, the authors found $Z_\psi -1= 1-Z_g$, presumably because of a subtle sign mistake somewhere.  They also concluded, erroneously we believe, that $g$ has a beta function.
 }

 \vskip 0.1in

 The perturbative corrections to the $\phi$ propagator vanish trivially, because they all involve fermion loops.  All fermion loops in this model vanish because of the cyclic product of theta functions.

 \vskip 0.1in

The beta function for $g$ can be computed by taking a derivative of
\begin{equation}
g_0 Z_\psi Z_\phi^{1/2} = g Z_g \mu^{\epsilon/2}
\end{equation}
with respect to the scale $\mu$.  
Since $Z_\phi = 1$ and $Z_g = Z_\phi$, $g$ is renormalized only by the classical contribution $\mu^{\epsilon/2}$ and is in fact marginal in $d=4$.  
We have only demonstrated $Z_g = Z_\psi$ up to one loop and without the Ward identity argument, one could worry there are differences at two loops.  Let us see how the two loop corrections validate the Ward identity next.

\subsection{Two Loop Corrections}
\begin{figure}
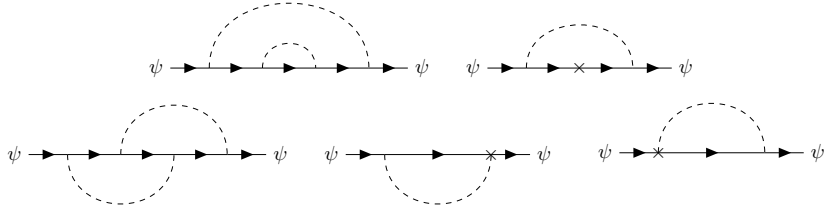

\begin{subfigure}{1.\textwidth}
\centering
\scalebox{0.7}{\proptwoloopone}
\scalebox{0.7}{\proptwolooponeCT}
\end{subfigure}
\begin{subfigure}{1.0\textwidth}
\centering
\scalebox{0.7}{\proptwolooptwo}
\scalebox{0.7}{\proptwolooptwoCTa}
\raisebox{1.8em}{
\scalebox{0.7}{\proptwolooptwoCTb}
}
\end{subfigure}
\caption{Two loop self energy and counter terms.}
\label{fig:proptwoloop}
\end{figure}

\begin{figure}
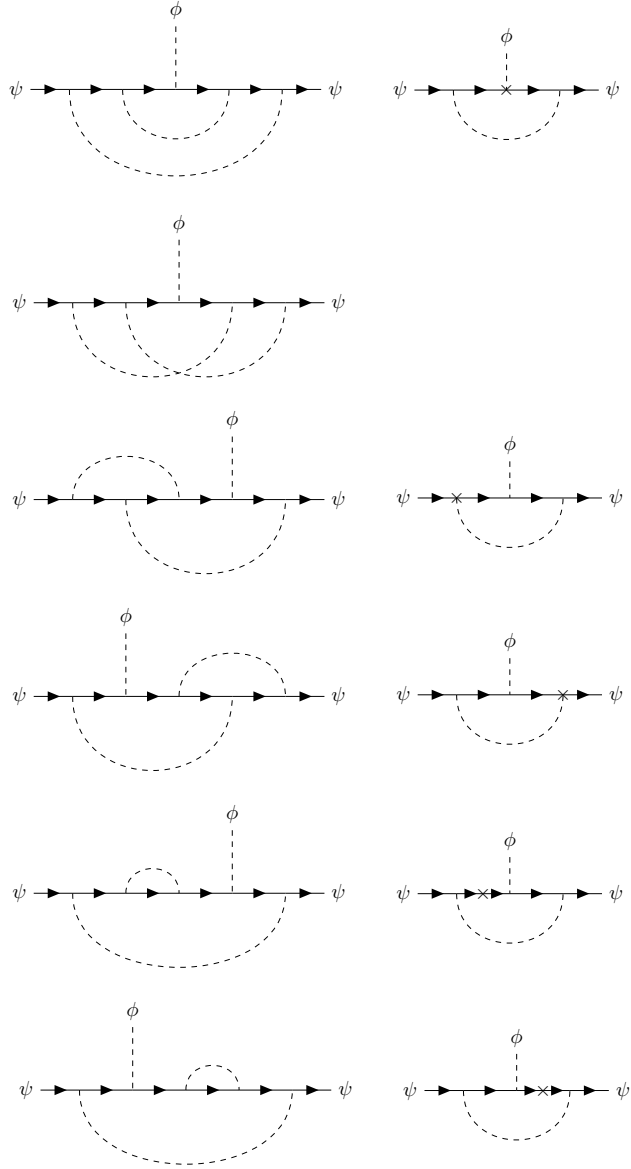

\hspace{3cm}
\begin{subfigure}{1.0\textwidth}
\scalebox{0.7}{\vertextwoloopone}
\scalebox{0.7}{
\raisebox{2.4em}{
\vertextwolooponeCT
}
}
\end{subfigure}
\begin{subfigure}{1.0\textwidth}
\hspace{3cm}
\scalebox{0.7}{\vertextwolooptwo}
\end{subfigure}
\begin{subfigure}{1.0\textwidth}
\hspace{3cm}
\scalebox{0.7}{\vertextwoloopthree}
\scalebox{0.7}{
\raisebox{1.7em}{
\vertextwoloopthreeCT
}
}
\end{subfigure}
\begin{subfigure}{1.0\textwidth}
\hspace{3cm}
\scalebox{0.7}{\vertextwoloopfour}
\scalebox{0.7}{
\raisebox{1.7em}{
\vertextwoloopfourCT
}
}
\end{subfigure}
\begin{subfigure}{1.0\textwidth}
\hspace{3cm}
\scalebox{0.7}{\vertextwoloopfive}
\scalebox{0.7}{
\raisebox{1.6em}{
\vertextwoloopfiveCT
}
}
\end{subfigure}
\begin{subfigure}{1.0\textwidth}
\hspace{3cm}
\scalebox{0.7}{
\vertextwoloopsix
}
\scalebox{0.7}{
\raisebox{1.6em}{
\vertextwoloopsixCT
}
}
\end{subfigure}
\caption{Two loop vertex diagrams and counter terms.}
\label{fig:vertextwoloop}
\end{figure}

There are two 1PI diagrams that contribute at $O(g^4)$ to the fermion propagator.
We invoke the shorthand $dp = d \omega d^{d-1} p_\perp / (2\pi)^d$.  
\begin{eqnarray}
i\Pi_{2,1}(\mu) &=& \int dp_1 \, dp_2 \frac{ (ig)^4 (-i)^2 (i)^3}{p_1^2 p_2^2
(\omega_1 + \mu)^2 (\omega_1 + \omega_2 + \mu)} \ , \\
i\Pi_{2,2}(\mu) &=& \int dp_1 \, dp_2 \frac{ (ig)^4 (-i)^2 (i)^3 }{p_1^2 p_2^2
(\omega_1 + \mu) (\omega_2 + \mu) (\omega_1 + \omega_2 + \mu)}  \ .
\end{eqnarray}
It is straightforward to carry out these integrals, integrating first over both $\omega_i$ and then over both $p_{i \perp}$. 
Expanding out near $d = 4 - \epsilon$ gives
\begin{eqnarray}
i\Pi_{2,1}(\mu) &=& -\frac{i \mu g^4}{32 \pi^4} \left( \frac{ 1}{ \epsilon^2} + \frac{1}{ \epsilon} \left( 1 + \log \frac{4 \pi}{\mu^2} + \psi \left( \frac{3}{2} \right) \right) + \ldots \right) \ ,
\\
i\Pi_{2,2}(\mu) &=& \frac{i \mu g^4}{16 \pi^4}  \left( \frac{ 1}{ \epsilon^2} + \frac{1}{ \epsilon} \left( \frac{1}{2} +\log \frac{4 \pi}{\mu^2} + \psi \left( \frac{3}{2} \right) \right) + \ldots \right) \ .
\end{eqnarray}
$\psi(x)$ is a poly gamma function.  Note $\psi(3/2) = 2-\gamma-\log(4)$.  
Note further that
\begin{eqnarray}
i \Pi_{2}(\mu) &\equiv& i \Pi_{2,1}(\mu) + i \Pi_{2,2}(\mu)  \\
&=& \frac{i \mu g^4 }{32 \pi^4} \left( \frac{1}{ \epsilon^2} + \frac{1}{\epsilon} \left( \log \frac{4 \pi}{\mu^2} + \psi \left(\frac{3}{2} \right) \right) + \ldots  \right) \ .\nonumber
\end{eqnarray}

\vskip 0.1in

There are three counter-term diagrams at one loop that we need to add to this result for $\Pi_2$.
Adding the appropriate counter terms to $i \Pi_{2,i}(\mu)$ individually, we get
\begin{eqnarray}
i \tilde \Pi_{2,1} &=& i \Pi_{2,1}(\mu) + (Z_\psi-1)  i \Pi_1(\mu) = \frac{ig^4 \mu}{32 \pi^4 \epsilon^2} - \frac{ ig^4 \mu}{32 \pi^4 \epsilon} + O(1) \ , \\
i \tilde \Pi_{2,2}  &=&   i \Pi_{2,1}(\mu) +2 (Z_g Z_\psi - 1) i \Pi_1(\mu)  =- \frac{ig^4 \mu}{16 \pi^4 \epsilon^2} + \frac{i g^4 \mu}{32 \pi^4 \epsilon} + O(1) \ .
\end{eqnarray}
Altogether, the result is
\[
i \tilde \Pi_2 = -\frac{i \mu g^4}{32 \pi^4 \epsilon^2} + O(1)  \ .
\]
The remaining divergence is free of $\mu$ dependence, aside from the overall factor of $\mu$, as it should be.  Also, we are finding no
anomalous dimension at $O(g^4)$.  

\vskip 0.1in

The vertex diagrams are less straightforward to carry out.  There are six
\begin{eqnarray}
iV_{2,1}(\mu_1, \mu_2) &=& \int\frac{(i g)^5 (-i)^2  dp_1 \, dp_2 }{p_1^2 p_2^2 (\omega_1 + \mu_1)(\omega_1+\mu_2)(\omega_1+\omega_2+ \mu_1)(\omega_1 + \omega_2 + \mu_2)} \ ,
\\
iV_{2,2}(\mu_1, \mu_2) &=& \int  \frac{(i g)^5 (-i)^2 dp_1 \, dp_2}{p_1^2 p_2^2 (\omega_1 + \mu_1)(\omega_2+\mu_2)(\omega_1+\omega_2+ \mu_1)(\omega_1 + \omega_2 + \mu_2)} \ ,
\\
iV_{2,3}(\mu_1, \mu_2) &=& \int  \frac{(i g)^5 (-i)^2 dp_1 \, dp_2}{p_1^2 p_2^2 (\omega_1 + \mu_1)(\omega_1+\mu_2)(\omega_1+\omega_2+ \mu_2)( \omega_2 + \mu_2)} \ ,
\\
iV_{2,4}(\mu_1, \mu_2) &=& \int  \frac{(i g)^5 (-i)^2 dp_1 \, dp_2}{p_1^2 p_2^2 (\omega_2 + \mu_1)(\omega_2+\mu_2)(\omega_1+\omega_2+ \mu_1)(\omega_1 + \mu_1)}
\ , \\
iV_{2,5}(\mu_1, \mu_2) &=& \int  \frac{(i g)^5 (-i)^2 dp_1 \, dp_2}{p_1^2 p_2^2 (\omega_1 + \mu_1)^2(\omega_1+\mu_2)(\omega_1+\omega_2+ \mu_1)} \ ,
\\
iV_{2,6}(\mu_1, \mu_2) &=& \int  \frac{(i g)^5 (-i)^2 dp_1 \, dp_2}{p_1^2 p_2^2 (\omega_1 + \mu_2)^2(\omega_1+\mu_1)(\omega_1+\omega_2+ \mu_2)} \ .
\end{eqnarray}
The integrals can be done by first integrating over the $\omega_i$ and then the $p_i$ although it helped to work with two nominally different dimensions $d_1$ and $d_2$ for the two $p_i$ integrals before setting them equal at the end.  The result is
\begin{eqnarray}
i V_{2,1}(\mu_1, \mu_2) &=& \frac{i g^5}{32 \pi^4} \left( \frac{1}{\epsilon^2} + \frac{1}{ \epsilon} \left( 1 - \gamma + \log \pi - \frac{ \mu_1 \log \mu_1^2 -  \mu_2 \log \mu_2^2}{\mu_1 - \mu_2} \right) \right) + \ldots
\ , \\
i V_{2,2}(\mu_1, \mu_2) &=& \frac{ig^5}{32 \pi^4} \frac{1}{\epsilon} + \ldots
\ , \\
i V_{2,3}(\mu_1, \mu_2) &=& \frac{ig^5}{32 \pi^4}  \left( \frac{1}{\epsilon^2} + \frac{1}{\epsilon} \left(2 - \gamma+ \log \pi -  \frac{ \mu_1 \log \mu_1^2 -  \mu_2 \log \mu_2^2}{\mu_1 - \mu_2} \right) \right) + \ldots
\ , \\
i V_{2,4}(\mu_1, \mu_2) &=&  \frac{ig^5}{32 \pi^4}  \left( \frac{1}{\epsilon^2} + \frac{1}{\epsilon} \left(2 - \gamma+ \log \pi -  \frac{ \mu_1 \log \mu_1^2 - \mu_2 \log \mu_2^2 }{\mu_1 - \mu_2} \right) \right) + \ldots
\ , \\
i V_{2,5}(\mu_1, \mu_2) &=& \frac{ig^5}{32 \pi^4}  \left(- \frac{1}{\epsilon^2} + \frac{1}{\epsilon} \left( -2 + \gamma - \log \pi 
+ \frac{ \mu_1 \log \mu_1^2 -  \mu_2 \log \mu_2^2}{\mu_1 - \mu_2}\right) \right) + \ldots
\ , \\
i V_{2,6}(\mu_1, \mu_2) &=& \frac{ig^5}{32 \pi^4}  \left(- \frac{1}{\epsilon^2}  + \frac{1}{\epsilon} \left( -2 + \gamma - \log \pi 
+ \frac{\mu_1 \log \mu_1^2 -  \mu_2 \log \mu_2^2}{\mu_1 - \mu_2}\right) \right) + \ldots
\ .
\end{eqnarray}
Diagrams 1, 3, 4, 5, and 6 all have a corresponding counter term diagram that needs to be subtracted.  
Individually, we find
\begin{eqnarray}
i \tilde V_{2,1} &=& i V_{2,1} + (Z_g - 1) i V_1 = -\frac{i g^5}{32 \pi^4 \epsilon^2} - \frac{ i g^5}{32 \pi^4 \epsilon} + O(1) \ , \\
i \tilde V_{2,3} &=& i V_{2,3} + (Z_g - 1) i V_1 = -\frac{ i g^5}{32 \pi^4 \epsilon^2} + O(1) \ , \\
i \tilde V_{2,4} &=&  i V_{2,4} + (Z_g  - 1) i V_1 = -\frac{ i g^5}{32 \pi^4 \epsilon^2} + O(1) \ , \\
i \tilde V_{2,5} &=& i V_{2,5} + (Z_\psi - 1) iV_1 =  \frac{ i g^5}{32 \pi^4 \epsilon^2}  + O(1) \ , \\
i \tilde V_{2,6} &=& i V_{2,6} + (Z_\psi - 1) iV_1 = \frac{ i g^5}{32 \pi^4 \epsilon^2} + O(1) \ .
\end{eqnarray}
Altogether, the result is
\[
i \tilde V_2 = -\frac{ ig^5}{32 \pi^4} \left( \frac{ 1}{\epsilon^2}  + O(1) \right) \ .
\]
The remaining divergences are free of $\mu_i$ dependence, as it should be.
One finds then equality of $Z_\psi$ and $Z_g$ at two loops:
\begin{eqnarray}
Z_\psi = Z_g &=& 1 - \frac{g^2}{4 \pi^2 \epsilon} + \frac{g^4}{32 \pi^4} \frac{1}{ \epsilon^2}  + O(g^6) + \ldots \ .
\end{eqnarray}
The fact that the $1/\epsilon^2$ terms match is required and a consistency check. 
The absence of a $O(g^4)/\epsilon$ piece means that the beta function for $g$ will be zero at two loops as well in $d=4$ dimensions.  One is just left with the tree level running in $4-\epsilon$ dimensions.
Indeed, the Ward identity guaranteed $Z_\psi = Z_g$ (and hence that $\beta=0$ at all loops in $d=4$), but it is nice to see these procedures confirming each other.  
The difficulty of calculating these Feynman diagrams in such a simple model also serves as a warning that while they may be good to fall back on when other options do not exist, Feynman diagrams are not always the most efficient way of proceeding.



\begin{thebibliography}{99}

\bibitem{Schwinger:1962tp}
J.~S.~Schwinger,
``Gauge Invariance and Mass. 2.,''
Phys. Rev. \textbf{128}, 2425-2429 (1962)

\bibitem{Fraser-Taliente:2024lea}
L.~Fraser-Taliente, C.~P.~Herzog and A.~Shrestha,
``A nonlocal Schwinger model,''
JHEP \textbf{06}, 252 (2025)
[\arxiv{2412.02514} [hep-th]].

\bibitem{Lauria:2020emq}
E.~Lauria, P.~Liendo, B.~C.~Van Rees and X.~Zhao,
``Line and surface defects for the free scalar field,''
JHEP \textbf{01} (2021), 060
[\arxiv{2005.02413} [hep-th]].

\bibitem{Herzog:2022jqv}
C.~P.~Herzog and A.~Shrestha,
``Conformal surface defects in Maxwell theory are trivial,''
JHEP \textbf{08} (2022), 282
[\arxiv{2202.09180} [hep-th]].

\bibitem{Behan:2020nsf}
C.~Behan, L.~Di Pietro, E.~Lauria and B.~C.~Van Rees,
``Bootstrapping boundary-localized interactions,''
JHEP \textbf{12}, 182 (2020)
[\arxiv{2009.03336} [hep-th]].

\bibitem{Behan:2021tcn}
C.~Behan, L.~Di Pietro, E.~Lauria and B.~C.~van Rees,
``Bootstrapping boundary-localized interactions II. Minimal models at the boundary,''
JHEP \textbf{03}, 146 (2022)
[\arxiv{2111.04747} [hep-th]].

\bibitem{DiPietro:2019hqe}
L.~Di Pietro, D.~Gaiotto, E.~Lauria and J.~Wu,
``3d Abelian Gauge Theories at the Boundary,''
JHEP \textbf{05}, 091 (2019)
[\arxiv{1902.09567} [hep-th]].



\bibitem{Rychkov:2016iqz}
S.~Rychkov,
``EPFL Lectures on Conformal Field Theory in D{$\geq$}3 Dimensions,''
[\arxiv{1601.05000} [hep-th]].


\bibitem{Billo:2016cpy}
M.~Bill{\`o}, V.~Gon{\c{c}}alves, E.~Lauria and M.~Meineri,
``Defects in conformal field theory,''
JHEP \textbf{04} (2016), 091
[\arxiv{1601.02883} [hep-th]].

\bibitem{Costa:2011mg}
M.~S.~Costa, J.~Penedones, D.~Poland and S.~Rychkov,
``Spinning Conformal Correlators,''
JHEP \textbf{11} (2011), 071
[\arxiv{1107.3554} [hep-th]].



\bibitem{McAvity:1995zd}
D.~M.~McAvity and H.~Osborn,
``Conformal field theories near a boundary in general dimensions,''
Nucl. Phys. B \textbf{455} (1995), 522-576
[\arxiv{cond-mat/9505127} [cond-mat]].

\bibitem{Dolan:2000ut}
F.~A.~Dolan and H.~Osborn,
``Conformal four point functions and the operator product expansion,''
Nucl. Phys. B \textbf{599}, 459-496 (2001)
[\arxiv{hep-th/0011040} [hep-th]].

\bibitem{Allais:2014fqa}
A.~Allais and S.~Sachdev,
``Spectral function of a localized fermion coupled to the Wilson-Fisher conformal field theory,''
Phys. Rev. B \textbf{90}, no.3, 035131 (2014)
[\arxiv{1406.3022} [cond-mat.str-el]].

\bibitem{Bashmakov:2024suh}
V.~Bashmakov and J.~Sisti,
``Exploring defects with degrees of freedom in free scalar CFTs,''
JHEP \textbf{03}, 147 (2025)
[\arxiv{2410.01716} [hep-th]].

\bibitem{Cuomo:2022xgw}
G.~Cuomo, Z.~Komargodski, M.~Mezei and A.~Raviv-Moshe,
``Spin impurities, Wilson lines and semiclassics,''
JHEP \textbf{06}, 112 (2022)
[\arxiv{2202.00040} [hep-th]].

\bibitem{Komargodski:2025jbu}
Z.~Komargodski, F.~K.~Popov and B.~C.~Rayhaun,
``Defect Anomalies, a Spin-Flux Duality, and Boson-Kondo Problems,''
[\arxiv{2508.14963} [hep-th]].

\bibitem{Vojta1999}
M.~Vojta, C.~Buragohain, and S.~Sachdev,
``Quantum impurity dynamics in two-dimensional antiferromagnets and superconductors,''
Phys.\ Rev.\ B, {\bf 61}, 
15152 (2000)
[\arxiv{cond-mat/9912020}].

\bibitem{Anninos:2016klf}
D.~Anninos and G.~A.~Silva,
``Solvable Quantum Grassmann Matrices,''
J. Stat. Mech. \textbf{1704}, no.4, 043102 (2017)
[\arxiv{1612.03795} [hep-th]].

\bibitem{Cuomo:2024psk}
G.~Cuomo, Y.~C.~He and Z.~Komargodski,
``Impurities with a cusp: general theory and 3d Ising,''
JHEP \textbf{11}, 061 (2024)
[\arxiv{2406.10186} [hep-th]].

\bibitem{Hull:2023iny}
C.~M.~Hull,
``Magnetic charges for the graviton,''
JHEP \textbf{05}, 257 (2024)
[\arxiv{2310.18441} [hep-th]].



\bibitem{Drukker:2022pxk}
N.~Drukker, Z.~Kong and G.~Sakkas,
``Broken Global Symmetries and Defect Conformal Manifolds,''
Phys. Rev. Lett. \textbf{129}, no.20, 201603 (2022)
[\arxiv{2203.17157} [hep-th]].

\bibitem{Herzog:2023dop}
C.~P.~Herzog and V.~Schaub,
``Tilting space of boundary conformal field theories,''
Phys. Rev. D \textbf{109}, no.6, L061701 (2024)
[\arxiv{2301.10789} [hep-th]].



\bibitem{Brown:1979pq}
L.~S.~Brown,
``Dimensional Regularization of Composite Operators in Scalar Field Theory,''
Annals Phys. \textbf{126} (1980), 135

\bibitem{Paulos:2015jfa}
M.~F.~Paulos, S.~Rychkov, B.~C.~van Rees and B.~Zan,
``Conformal Invariance in the Long-Range Ising Model,''
Nucl. Phys. B \textbf{902} (2016), 246-291
[\arxiv{1509.00008} [hep-th]].

\bibitem{Ge:2025fsm}
D.~Ge and Y.~Nakayama,
``Non-Factorizing Interface in the Two-Dimensional Long-Range Ising Model,''
[\arxiv{2505.15018} [hep-th]].



\bibitem{Klebanov:2011gs}
I.~R.~Klebanov, S.~S.~Pufu and B.~R.~Safdi,
``F-Theorem without Supersymmetry,''
JHEP \textbf{10} (2011), 038
[\arxiv{1105.4598} [hep-th]].

\bibitem{Cuomo:2021kfm}
G.~Cuomo, Z.~Komargodski and M.~Mezei,
``Localized magnetic field in the O(N) model,''
JHEP \textbf{02} (2022), 134
[\arxiv{2112.10634} [hep-th]].

\bibitem{Ge:2024hei}
D.~Ge, T.~Nishioka and S.~Shimamori,
``Localized RG flows on composite defects and $ \mathcal{C} $-theorem,''
JHEP \textbf{02} (2025), 012
[\arxiv{2408.04428} [hep-th]].

\bibitem{Hogervorst:2015akt}
M.~Hogervorst, S.~Rychkov and B.~C.~van Rees,
``Unitarity violation at the Wilson-Fisher fixed point in 4-$\epsilon$ dimensions,''
Phys. Rev. D \textbf{93}, no.12, 125025 (2016)
[arXiv:1512.00013 [hep-th]].

\bibitem{Herzog:2020bqw}
C.~P.~Herzog and A.~Shrestha,
``Two point functions in defect CFTs,''
JHEP \textbf{04} (2021), 226
[\arxiv{2010.04995} [hep-th]].


\bibitem{Okuyama:2023fge}
Y.~Okuyama,
``Aspects of critical O$(N)$ model with boundary and defect,''
[\arxiv{2401.15336} [hep-th]].

\bibitem{Marchetto:2023fcw}
E.~Marchetto, A.~Miscioscia and E.~Pomoni,
``Broken (super) conformal Ward identities at finite temperature,''
JHEP \textbf{12} (2023), 186
[\arxiv{2306.12417} [hep-th]].

\bibitem{Barrat:2024aoa}
J.~Barrat, B.~Fiol, E.~Marchetto, A.~Miscioscia and E.~Pomoni,
``Conformal line defects at finite temperature,''
SciPost Phys. \textbf{18} (2025) no.1, 018
[\arxiv{2407.14600} [hep-th]].

\bibitem{Levine:2023ywq}
N.~Levine and M.~F.~Paulos,
``Bootstrapping bulk locality. Part I: Sum rules for AdS form factors,''
JHEP \textbf{01}, 049 (2024)
doi:10.1007/JHEP01(2024)049
[\arxiv{2305.07078} [hep-th]].

\bibitem{Levine:2024wqn}
N.~Levine and M.~F.~Paulos,
``Bootstrapping bulk locality. Part II: Interacting functionals,''
[\arxiv{2408.00572} [hep-th]].

\bibitem{Loparco:2025aag}
M.~Loparco, G.~Mathys, J.~Penedones, J.~Qiao and X.~Zhao,
``Locality constraints in AdS$_2$ without parity,''
[\arxiv{2511.20749} [hep-th]].


\end{thebibliography}
\end{document}